\documentclass[fleqn,usenatbib]{mnras}

\usepackage{newtxtext,newtxmath}

\usepackage[T1]{fontenc}

\DeclareRobustCommand{\VAN}[3]{#2}
\let\VANthebibliography\thebibliography
\def\thebibliography{\DeclareRobustCommand{\VAN}[3]{##3}\VANthebibliography}

\usepackage{graphicx}	
\usepackage{amsmath}	
\usepackage{multirow}
\usepackage{hyperref}
\hypersetup{colorlinks=true}

\title[SLSN~I cosmology]{The first Hubble diagram and cosmological constraints using superluminous supernovae}

\author[C. Inserra et al.]{C. Inserra$^{1}$\thanks{InserraC@cardiff.ac.uk (CI)},
M.~Sullivan$^{2}$, 
C.~R. Angus$^{3}$, 
E.~Macaulay$^{4}$, 
R.~C.~Nichol$^{5}$, 
M.~Smith$^{2,58}$,
\newauthor 
C.~Frohmaier$^{5}$, C.~P.~Guti\'errez$^{2,59,60}$, M. Vicenzi$^{5}$,       
A.~M\"oller$^{6}$,  D.~Brout$^{7}$, P.~J.~Brown$^{8}$, 
\newauthor T.~M.~Davis$^{9}$, 
 C.~B.~D'Andrea$^{7}$,  L.~Galbany$^{10}$, R.~Kessler$^{11,12}$, A.~G.~Kim$^{13}$, Y.-C.~Pan$^{14}$, \newauthor M.~Pursiainen$^{2}$, 
 D.~Scolnic$^{12}$, B.~P.~Thomas$^{5}$,  P.~Wiseman$^{2}$, 
 T.~M.~C.~Abbott$^{15}$, J.~Annis$^{16}$,  \newauthor S.~Avila$^{17}$, E.~Bertin$^{18,19}$,D.~Brooks$^{20}$, D.~L.~Burke$^{21,22}$, A.~Carnero~Rosell$^{23,24}$,  \newauthor M.~Carrasco~Kind $^{25,26}$, 
J.~Carretero$^{27}$, F.~J.~Castander$^{28,29}$, R.~Cawthon$^{30}$,  
  S.~Desai$^{31}$, \newauthor H.~T.~Diehl$^{16}$, T.~F.~Eifler$^{32,33}$, D.~A.~Finley$^{16}$, B.~Flaugher$^{16}$, P.~Fosalba$^{28,29}$, J.~Frieman$^{12,16}$,\newauthor
  J.~Garcia-Bellido$^{17}$, E.~Gaztanaga$^{28,29}$, D.~W.~Gerdes$^{34,35}$, T.~Giannantonio$^{36,37}$,  \newauthor D.~Gruen$^{21,22,38}$, R.~A.~Gruendl$^{25,26}$,
 J.~Gschwend$^{24,39}$, G.~Gutierrez$^{16}$, D.~L.~Hollowood$^{40}$, \newauthor K.~Honscheid$^{41}$, D.~J.~James$^{42}$, E.~Krause$^{32}$, K.~Kuehn$^{43}$, N.~Kuropatkin$^{16}$, T.~S.~Li$^{11,16}$,\newauthor C.~Lidman$^{44}$, M.~Lima$^{24,45}$, M.~A.~G.~Maia$^{24,39}$, J.~L.~Marshall$^{7}$,
 P.~Martini$^{41}$, F.~Menanteau$^{25,26}$, \newauthor R.~Miquel$^{27,46}$, A.~A.~Plazas~Malag\'on$^{47}$, A.~K.~Romer$^{48}$, A.~Roodman$^{21,22}$, M.~Sako$^{7}$, \newauthor E.~Sanchez$^{23}$, V.~Scarpine$^{16}$, M.~Schubnell$^{35}$,
 S.~Serrano$^{28,29}$, I.~Sevilla-Noarbe$^{23}$, \newauthor M. Soares-Santos$^{49}$, F.~Sobreira$^{23,50}$, E.~Suchyta$^{51}$, M.~E.~C.~Swanson$^{26}$, G.~Tarle$^{35}$,\newauthor
 D.~Thomas$^{5}$, D.~L.~Tucker$^{16}$, V.~Vikram$^{52}$, A.~R.~Walker$^{15}$, Y.~Zhang$^{16}$,
 J.~Asorey$^{53}$,\newauthor J.~Calcino$^{54}$, D.~Carollo$^{55}$,  K.~Glazebrook$^{56}$,  S.~R.~Hinton$^{54}$, J.~K.~Hoormann$^{54}$, G.~F.~Lewis$^{57}$, \newauthor
 R.~Sharp$^{44}$, E.~Swann$^{2}$ and B.~E.~Tucker$^{44}$. 
 \newauthor\hspace{7cm} (DES Collaboration)
\\
\emph{\normalsize Affiliations are listed at the end of the paper}
}
\date{Accepted XXX. Received YYY; in original form ZZZ}

\pubyear{2020}

\begin{document}
\label{firstpage}
\pagerange{\pageref{firstpage}--\pageref{lastpage}}
\maketitle

\begin{abstract}
We present the first Hubble diagram of superluminous supernovae (SLSNe) out to a redshift of two, together with constraints on the matter density, $\Omega_{\rm M}$, and the dark energy equation-of-state parameter, $w(\equiv p/\rho)$. We build a sample of 20 cosmologically useful SLSNe~I based on light curve and spectroscopy quality cuts. We confirm the robustness of the peak decline SLSN~I standardization relation with a larger dataset and improved fitting techniques than previous works. We then solve the SLSN model based on the above standardisation via minimisation of the $\chi^2$ computed from a covariance matrix which includes statistical and systematic uncertainties.
For a spatially flat $\Lambda$CDM cosmological model, we find $\Omega_{\rm M}=0.38^{+0.24}_{-0.19}$, with a rms of 0.27 mag for the residuals of the distance moduli. For a $w_0w_a$CDM  cosmological model, the addition of SLSNe~I to a `baseline' measurement consisting of Planck temperature  together with type Ia supernovae, results in a small improvement in the constraints of $w_0$ and $w_a$ of 4\%. We present simulations of future surveys with 868 and 492 SLSNe I (depending on the configuration used) and show that such a sample can deliver cosmological constraints in a flat $\Lambda$CDM model with the same precision (considering only statistical uncertainties) as current surveys that use type Ia supernovae, while providing a factor 2-3 improvement in the precision of the constraints on the time variation of dark energy, $w_0$ and $w_a$.
This paper represents the proof-of-concept for superluminous supernova cosmology, and demonstrates they can provide an independent test of cosmology in the high-redshift ($z>1$) universe.
\end{abstract}




\section{Introduction}

Twenty years have passed since observations of type Ia supernovae (SNe Ia) provided the first direct evidence for cosmic acceleration \citep{riess98,perlmutter99}. The physical origin of this acceleration is unknown, but is often described by a phenomenon called `dark energy'.
Combining SN Ia observations with measurements of large-scale structure \citep[e.g.][]{percival07,bao_sdss14} and the cosmic microwave background \citep[CMB, e.g.][]{wmap,planck16} shows that dark energy is the major component ($\approx$70\%) of the energy density of the Universe at the present epoch. SNe Ia present a direct and mature method of probing this dark energy via its equation-of-state parameter $w$. Current SN-only measurements provide a precision of 20\%, dropping to 4-5\% when combined with measurements of the CMB \citep{scolnic2018,descosmo}. However, SNe Ia at $z\gtrapprox1.2$ are exceptionally challenging to observe from the ground, and thus assembling large samples at these high redshifts is very time-consuming \citep{riess18} due to both the faintness of SNe Ia and line-blanketing in their ultraviolet spectra.

Hydrogen-free superluminous supernovae \citep[SLSNe I,][]{qu11,gy12} are significantly more luminous, do not suffer the same degree of line-blanketing as SNe Ia and have been observed at higher redshifts than SNe Ia, photometrically out to $z\sim4$ \citep{cooke} and spectroscopically out to $z\sim2$ \citep{mat18}.
These objects are characterised by a distinctive spectroscopic evolution linking them with massive stars \citep{pasto10}, and show remarkable peak luminosities $\bar{M}<-21$\,mag \citep{lu17,in18c,decia18,angus18}. Their light curve decline rates and colour evolution are similar,  suggesting these events may be standardizable \citep{in14} via a peak-decline relation in a synthetic band centered at 400 nm. 

\section{Superluminous supernova data sample}\label{sec:sample}
\subsection{The superluminous supernova definition and subtypes}
The challenge in using SLSNe~I as standardisable candles is to find a robust definition of the class that does not simply depend on their luminosity and, ideally, an association with a common explosion mechanism and progenitor scenario to decrease contamination. 

In the previous work about SLSNe~I standardization \citep{in14} two observational subclasses of SLSNe~I were used and, at the time, it was not immediately clear if these were distinct or if there was a continuum of properties bridging the gap between them. However, this distinction is important if they are to be utilised as standardizable candles, since the bulk of the population (and those showing the strongest correlation parameters) are SLSNe~I with light curve evolution similar to SN2010gx \citep[][hereafter referred to as Fast]{pasto10}. The other subtype, encompassing objects similar to SN2007bi \citep[][called Slow hereafter]{gy09,young10}, instead increases the scatter on the proposed correlations. This increase in the scatter may be due to the presence of interaction in these objects, which is observed in light curves and spectra \citep[e.g.][]{yan15,ni16,in17}. 
More recent works have shown that a distinct division can indeed be made between these two classes \citep{in18c,quimby18,galyam19,in19}, Fast (F) and Slow (S).  This distinction is possible via light curve, from peak to +30 days, and spectra information, at roughly +10 days and up to +30 days. This classification, based on K-means partitional cluster analysis, also requires photospheric velocity information derived from the Fe~{\sc ii} $\lambda$5169 line. Those evolving more slowly frequently show signatures of an interaction with a circumstellar medium \citep{yan17b,ni16,in17,in19}, perhaps pointing to a different progenitor scenario. The first step in building a homogeneous sample is then to define what is a cosmologically useful SLSN~I based on its spectrophotometric behaviour.

\begin{figure*}
\center

\includegraphics[width=16cm]{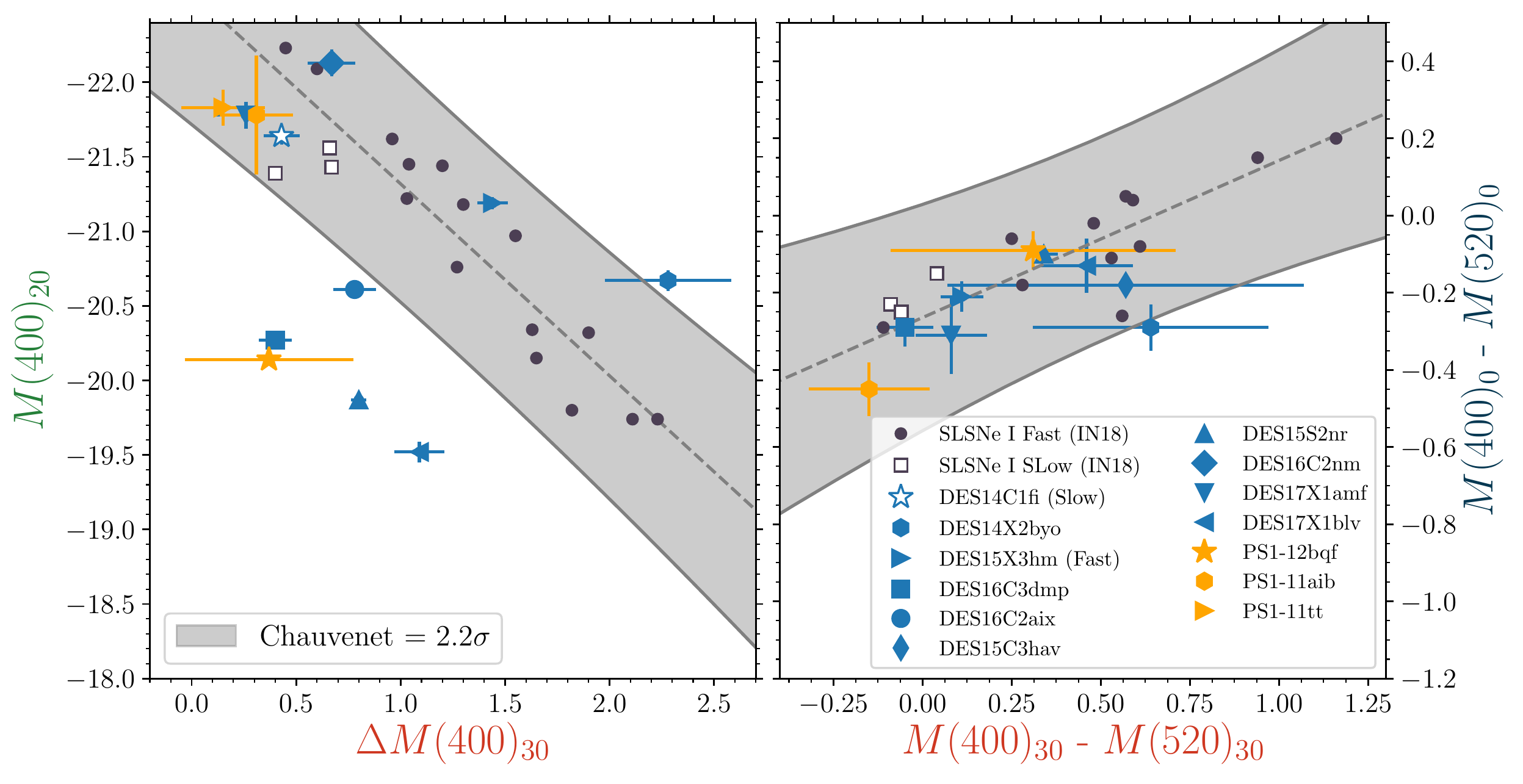}
\caption{Our third criterion for SLSN~I selection. A reduced and modified version of the Four Observables Parameter Space (4OPS) for SLSNe~I. Data are taken from DES, literature, and other surveys' sample papers, that made our first two quality cuts. The left panel shows the magnitude at 20 days post peak, vs the decline rate over 30 days past peak. We have used this relationship to replace the peak decline relation ($M(400)_{\rm 0}$ vs $\Delta M(400)_{\rm 30}$), which had been used in the original 4OPS paper. The right panel shows the colour at peak vs the colour at 30 days post peak. The literature objects, both Fast (circles) and Slow (open squares) are shown together to their best fit of the weighted linear regression (dashed, black line) and with the $2.2\sigma$ confidence bands defined by the Chauvenet's criterion. We also include DES16C2nm, which was not presented before in the literature sample \citep{in18c}. }
\label{fig:4OPS}
\end{figure*}

\begin{table*}
\caption{SLSN~I complete sample\label{table:data}. Literature SLSNe~I have been broken down in terms of single object papers, for which the contribution from each survey to the total is reported between parentheses, and big sample papers of previously unpublished objects.}
\begin{center}
\resizebox{18cm}{!}{%
\begin{tabular}{lcccc}
\hline
\hline
Sample source &  SLSN~I candidates & light curve and spectra (quality cuts 1 and 2) & 4OPS (quality cut 3)& Reference\\
\hline
Literature (4 DES - 6 PS1 - 11 PTF/iPTF) & 40 & 23 & 20& \citep{in18c}\\
\hspace{1.5cm} - PS1 (Medium Deep survey) & 9 & 3 &2& \citep{lu17}\\
\hspace{1.5cm} - PTF/iPTF & 15 & 0 & 0 & \citep{decia18}\\
DES & 17 & 9 & 3 & \citep{angus18}\\
\hline
\end{tabular}%
}
\end{center}
\end{table*}

\subsection{The sample construction}
We begin to build our sample with all spectroscopically-confirmed SLSNe~I available in the literature, starting with the 40 SLSNe~I from the compilation of \citet{in18c}, and adding 9 from PanSTARRS-1 \citep{lu17}, 15 from the Palomar Transient Factory (PTF) and intermediate PTF (iPTF) \citep{decia18}, and 17 from the Dark Energy Survey \citep[DES,][]{angus18}. 

We apply light-curve quality cuts to our sample in order to assemble a sub-sample with adequate photometric coverage in a synthetic rest-frame filter centered at 400\,nm,  which was previously used to test their standardisation \citep{in14}.
To fulfill our quality cuts, the objects need a light curve covering -15 to +30 days in the rest-frame and without multiple peaks to remove ambiguity in identifying the main peak and thus measuring phases (first quality cut). These requirements do not exclude the presence of early time `bumps' \citep{ni15a,smi16}. Furthermore, a spectrum taken between $-15$ to $+30$ days rest-frame must also be available (second quality cut). 

The literature sample has 23 such SLSNe~I. We apply the same selection criteria to the DES SLSN~I candidates \citep{angus18}. Of 17 events, 10 passed the light-curve quality criteria, and all have at least one spectrum in the required phase range. This retention fraction of 58\% is somewhat higher than what seen in the literature sample and can be explained by the DES cadence during the six months observing season and higher redshift of several objects \citep{htd2,htd1,angus18}. 
Of the 9 additional, and previously unpublished, events within the Pan-STARRS1 (PS1) Medium Deep survey \citep{lu17} only 3 passed our quality cuts. We also examined events from the PTF/iPTF sample, but none had a sufficient sampling in the rest-frame 400nm.

However, these published SLSN samples are very heterogeneous in their target selection and therefore we apply a homogeneous method to select our objects. 
To do this, our third quality cut is based on a statistical approach to identify SLSNe~I from their multi-band photometric behaviour and the distribution of the candidates on the hyper-surface defined by four photometric variables \citep[4OPS,][]{in18c}. This parameter space uses the peak luminosity in the 400 nm filter, the decline in magnitudes in the 400 nm filter over the 30 days following peak brightness, the 400–520 color at peak  and the 400–520 color at +30 days. This hyper-surface provides information on the overall SLSN~I population evolution, and it is a valuable alternative to identifying SLSNe~I when only a single spectrum bearing resemblance to other SLSNe~I is available, as is the case for several SLSN~I candidates.  However, one of the key relationships describing this hyper-surface is the standardisation relation. To remove any potential bias in the way we select cosmologically useful SLSNe~I, we decide to use a slightly different hyper-plane than the original. This is needed because if we would have used the original hyper-surface, we would have implicitly selected SLSNe fitting the standardisation relation and then creating a circular argument.  relation.
The alterations made to the hyper-surface here used with respect to the original one \citep[][]{in18c} are the following: 1) we replace panel A \citep{in18c} with a different decline relation ($M(400)_{\rm 20}$ vs $\Delta M(400)_{\rm 30}$) to avoid introducing any biases over the fact that the peak-decline relation  ($M(400)_{\rm 0}$ vs $\Delta M(400)_{\rm 30}$) is a consequence of our selection criteria; 2) we present the decline panel with the colour panel, which is panel D of the original 4OPS. A version with $M(400)_{\rm 0}$ rather than $M(400)_{\rm 20}$ as y-axis of the left panel can be found in the Appendix~\ref{app:4OPS}.

We will refer to core SLSNe~I as those which lie within $2.2\sigma$ of the hyper-plane here constructed (see Figure~\ref{fig:4OPS}). The choice of 2.2$\sigma$ is driven by the Chauvenet's criterion value for a sample of this size. Chauvenet's criterion is an outlier detection method used in experimental physics, but it has also been applied to supernova cosmology \citep[e.g.][]{conley2011,betoule2014}, and can be used for Gaussian distributed datasets such as the one presented here \citep{in14}.
Moreover, each slice of the hyper-surface described by the 2.2$\sigma$ band contour 
spans $\sim3$ mag in the luminosity scale, supporting the fact that we are not excluding {\it a priori} the luminosity extremes of the population. In some cases, like DES14X2byo, the supernova does not make the core population due to a different colour evolution with respect to the SLSN~I core, i.e. missing the 2.2$\sigma$ region in the colour panel (see right panel of Figure~\ref{fig:4OPS}). We also note that all SLSNe~I candidates not making into the 4OPS bands show an early `bump' or undulation in the light curve after 30 days. Hence, we would have the same dataset if the third quality cut was linked to the morphology of the whole light curve instead of the statistical description. This further inspection supports the fact that our selection criteria are not driven by any underlying relations. All explored relations, fit parameters and statistical results of our sample are reported in Table~\ref{table:fit}.

The literature sample has 20 objects belonging to the core SLSN~I population. Of the 10 DES objects only 4 reside in the hyper-surface (of which one event is a Slow, see Figure~\ref{fig:4OPS}), while the others show a similar trend in their luminosity evolution but at lower luminosities, down to roughly {\it M}(400)$\sim -19.3$\,mag. This suggests a population of transients similar to SLSNe~I with peak magnitudes down to those of normal core-collapse SNe \citep{angus18}.
Only 2 of the PS1 objects lye in the core distribution. The events lying outside the core population have a similar photometric behaviour to the DES SLSNe~I that do not meet the selection criteria (see Figure~\ref{fig:4OPS}). 

Hence, the final sample fulfilling all criteria comprises 26 SLSNe~I (see Table~\ref{table:data}), of which 15 belong to the Fast subclass and 6 to the Slow subclass; the remaining 5, due to their high redshift that prevent measuring the Fe~{\sc ii} line, have insufficient spectroscopic data to assign a subclass and hence will be labelled as `No Subclass' (NS).
Indeed, it is impossible to observe the Fe~{\sc ii} line at $z\gtrsim0.8$ with optical spectroscopy, although infrared spectroscopy and/or an analysis based on the ejecta velocity of UV lines \citep[see][]{galyam19b} might extend this spectroscopic division out to $z\sim3.7$. 

Because of the presence of interaction in the Slow subclass \citep{yan17b,ni16,in17,in19}, which add an additional source of scatter, we use the Fast and those with No Subclass in our analysis, for a final sample of 20 SLSNe~I in total. There may, of course, be Slow subtypes present in the NS sample that may bias our results, but such contamination is unavoidable with the current dataset if we want to probe the high-redshift region ($z>1.5$) unexplored with SNe Ia.  The impact of this contamination is beyond the scope of this paper, which is intended as a proof of concept. Nevertheless, their addition will provide a less prominent contamination than previous studies in which both Fast and Slow events were included in the analysis.

\subsection{The reddening assumption}\label{ss:red}
Before attempting any cosmological analysis or standardisation relation, it is important that our assumption of negligible host galaxy reddening (or local reddening at the supernova location) is accurate. This assumption is supported by multiple factors. Firstly, the color distribution around peak for SLSNe~I is quite narrow in the optical and ultraviolet (UV) irrespective of redshift; 0.46 mag in the optical \citep{in14,in18c} and 0.53 mag in the UV \citep{mat18}. This scatter would be significantly increased in the case of environmental reddening. Secondly, SLSNe~I UV peak light distribution exhibits a small scatter ($\Delta$M$_{\rm UV}<1$ mag) regardless of the redshift \citep{mat18}, host reddening would have strongly affected the UV distribution causing a scatter larger than currently observed. Thirdly, SLSNe~I spectra around peak epoch show the temperature sensitive O~{\sc ii} lines (12000 $<$ T(K) $<$ 16000) \citep{qu13,maz16}, hence reddened sources would apparently lie outside of this temperature range. This is not observed. Finally, SLSNe~I explode in dwarf, metal poor galaxies similar to those hosting long Gamma-Ray Bursts \citep[e.g.][]{lu14} which have low dust content as shown by high-redshift galaxies hosting Gamma-Ray Bursts \citep{wiseman17}. Only a confirmed SLSN I and a candidate one, SN2017egm \citep{chen17} and SN2018don \citep{lunnan19}, both Slow events, show a significant host reddening (E(B-V)$>$0.2). We also exclude any dust formation in the material surrounding SLSNe~I (i.e. circum-stellar material) 
because circum-stellar material has only been indirectly observed (i.e. light-curve undulations) in the Slow type. Since SLSNe~I, intrinsically, are stripped envelope SNe boosted in luminosity \citep{pasto10,in13}, the distance and physical conditions of the material expelled by the star before undergoing its final demise are not favourable for early dust production causing a reddening excess of E(B-V) $>$ 0.02 at the timescale needed for our analysis (i.e. during the photospheric phase).

\section{Supernova light curves}\label{sec:lc}
To estimate and model the light curves of SLSNe~I around peak, ($-15\leq\mathrm{phase\, (days)}\leq30$) we first {\it k-}correct the observed magnitudes to the 400\,nm and 520\,nm synthetic filters with the \textsc{snake}\footnote{\url{https://github.com/cinserra/S3}} software package \citep{in18b}, which 
also estimates the uncertainties on the $k$-corrections. The apparent peak magnitude in rest-frame 400nm band ($m(400)$) is given by
\begin{equation}
m(400) = m(X) -   k_{X\rightarrow400}
\end{equation}
where $m(X)$ is the observed apparent magnitude in passband $X$ and the passband is chosen from the observed filters available for each SLSN to be closest in wavelength to 400nm after accounting for the cosmological redshift ($1+z$). This process is known as cross-filter {\it k-}correction \citep{kim96}. $k_{X\rightarrow400}$ is the {\it k-}correction from this passband to the 400nm passband. An analogous relation is used for the 520\,nm passband.
When observed spectra for a given SLSN~I are not available, we use an average SLSN~I time-series spectral energy distribution (SED) to compute the {\it k-}correction \citep{pr17}. We correct all our observed photometry for Milky Way extinction \citep{sf11}, but make no corrections for extinction in the SLSN~I host galaxies, which is assumed to be small \citep{ni15,le15,in19} as suggested by the small scatter in the colour distribution of SLSNe in the optical \citep{in18c,in19} and UV \citep {mat18} (see Section~\ref{ss:red}). We then use Gaussian processes (GPs) regression \citep{bishop,gps} to fit the light curves, using the Python package {\sc george} and a Matern-3/2 kernel to perform our GP regression of SLSN~I light curves and derive the uncertainties \citep[see][for a more in-depth description]{in18c} \footnote{\url{https://github.com/cinserra/Gaussian-Processes-GP-}}. 

\begin{table*}
\caption{SLSNe I and their peak-decline relation values, together with the quantities used in the third criterion approach \label{table:datarel}.}
\begin{center}
\resizebox{17cm}{!}{%
\begin{tabular}{lcccc|ccc}
\hline
\hline
ID & $z$ & SLSN~I type & $M$(400)$_{\rm 0}$ & $\Delta M$(400)$_{\rm 30}$ & $M$(400)$_{\rm 20}$ & $M(400)_{\rm 0}-M(520)_{\rm 0}$ & $M(400)_{\rm 30}-M(520)_{\rm 30}$\\
\hline
Gaia16apd & 0.102 & F & -21.87 (0.04) & 0.69 (0.06) & -21.18 (0.04) & -0.18 (0.07) & 0.28 (0.07) \\
SN2011ke & 0.143 & F & -21.23 (0.09) & 0.89 (0.09) & -20.34 (0.02) & 0.04 (0.13) & 0.59 (0.03)\\
SN2012il &0.175 & F & -21.54 (0.10) & 1.39 (0.17) &  -20.15 (0.14) & -0.02 (0.11) & 0.48 (0.13) \\
PTF11rks & 0.190 & F & -20.61 (0.05) & 0.87 (0.07) & -19.74 (0.05) & 0.20 (0.06)& 1.16 (0.15)\\
SN2010gx & 0.230 & F & -21.73 (0.02) & 0.76 (0.03) & -20.97 (0.02) & -0.11 (0.02) & 0.53 (0.03)\\
SN2011kf & 0.245 & F & -21.74 (0.15) & 0.52 (0.18) & -21.22 (0.02) & ... & ... \\
LSQ12dlf &0.255 & F & -21.52 (0.03) & 0.76 (0.04)& -20.76 (0.03) & 0.05 (0.03) &  0.57 (0.10)\\
LSQ14mo &0.256 & F & -21.04 (0.05) & 1.30 (0.14) & -19.74 (0.13) & -0.08 (0.04) & 0.61 (0.02) \\
PTF09cnd &0.258 & F & -22.16 (0.08) & 0.71 (0.14) & -21.45 (0.12) & ... & ... \\
SN2013dg &0.265 & F & -21.35 (0.05) & 1.03 (0.06) & -20.32 (0.03) & -0.26 (0.08) & 0.56 (0.10)\\
PS1-10bzj & 0.650 & F & -21.03 (0.06) & 1.23 (0.32) & -19.08 (0.31) & 0.15 (0.11) & 0.94 (0.25)\\
iPTF13ajg &0.740 & F & -22.42 (0.07) & 0.19 (0.10) & -22.23 (0.07) & -0.29 (0.09) & -0.11 (0.09) \\
DES15X3hm & 0.860 & F & -21.94 (0.06) & 1.44 (0.07) & -21.19 (0.04)  & -0.21 (0.04) & 0.11 (0.06) \\
DES17X1amf & 0.920 & NS & -21.97 (0.07) & 0.26 (0.15) & -21.78 (0.09) & -0.31 (0.10)& 0.08 (0.10) \\
PS1-10ky & 0.956 & F & -22.05 (0.06) & 0.61 (0.07) & -21.44 (0.04) & -0.06 (0.07) & 0.25 (0.06) \\
PS1-11aib &  0.997 & NS & -22.05 (0.07) & 0.31 (0.17) & -21.78 (0.40)  &-0.45 (0.07) & -0.15 (0.07) \\
SCP-06F6&1.189 & F & -22.19 (0.03) & 0.57 (0.15) & -21.62 (0.15) & ... & ... \\
PS1-11tt  &  1.283 & NS & -21.89 (0.16) & 0.15 (0.20) & -21.83 (0.12) & ... & ...\\
PS1-11bam&1.565 & NS & -22.45 (0.10) & 0.36 (0.14) & -22.09 (0.10) & ... & ... \\
DES16C2nm & 1.998 & NS & -22.52	(0.10) & 0.67 (0.11) & -22.13 (0.09) & ... & ... \\
\hline
\end{tabular}%
}
\end{center}
\end{table*}

\begin{table*}
\caption{Fit parameters and statistical results of our sample\label{table:fit}}
\begin{center}
\resizebox{17cm}{!}{%
\begin{tabular}{cccccccc}
\hline
\hline
$x$ & $y$ & SLSN~I type &N (objects)& $\beta$ & $\alpha$ & $\sigma$  & $\widetilde{\chi}^2$\\
\hline
\multirow{ 3}{*}{$\Delta M$(400)$_{\rm 30}$} & \multirow{3}{*}{$M$(400)$_{\rm 0}$} &F & 15 & $-23.09\pm0.28$& $1.01\pm0.17$ & $0.26\pm0.22$   &1.23\\
&& F+NS & 20 & $-22.62\pm0.19$ & $0.72\pm0.14$ & $0.33\pm0.22$ &  1.73\\
&& F+NS+S&25 &$-22.31\pm0.15$& $0.53\pm0.13$ & $0.36\pm0.22$ &  2.31 \\
\hline 
\multirow{3}{*}{$\Delta$($M$(400)$_{\rm 30}$ - $M$(520)$_{\rm 30}$)} & \multirow{3}{*}{$M$(400)$_{\rm 0}$}& F & 12 & $-22.76\pm0.35$& $2.18\pm0.62$ & $0.29\pm0.30$ &  1.45\\
&& F+NS & 14 &$-22.79\pm0.30$ & $2.22\pm0.55$ & $0.25\pm0.25$ &  1.20 \\
&& F+NS+S & 18 & $-22.31\pm0.20$& $1.55\pm0.41$ & $0.32\pm0.24$ & 1.91\\
\hline 
\multirow{3}{*}{$M$(400)$_{\rm 20}$ - $M$(520)$_{\rm 20}$} & \multirow{ 3}{*}{$M$(400)$_{\rm 0}$}& F & 12 & $-21.97\pm0.14$& $1.97\pm0.47$ & $0.31\pm0.27$ &  1.58 \\
&& F+NS & 14  & $-21.96\pm0.12$ & $1.89\pm0.41$ & $0.28\pm0.24$ &  1.41\\
&& F+NS+S& 18 & $-21.77\pm0.09$& $1.32\pm0.33$ & $0.33\pm0.24$ & 1.74 \\
\hline 
\multirow{3}{*}{$M$(400)$_{\rm 30}$ - $M$(520)$_{\rm 30}$} & \multirow{ 3}{*}{$M$(400)$_{\rm 0}$}& F & 12 & $-22.27\pm0.14$& $1.51\pm0.27$ & $0.21\pm0.19$ &  0.87\\
&& F+NS & 14 &$-22.21\pm0.11$ & $1.41\pm0.23$ & $0.19\pm0.17$ &  0.95\\
&& F+NS+S & 18 & $-21.95\pm0.09$& $1.02\pm0.21$ & $0.28\pm0.20$ &  1.52 \\
\hline 
\multirow{3}{*}{$\Delta M$(400)$_{\rm 30}$} & \multirow{ 3}{*}{$M$(400)$_{\rm 20}$}& F & 15 & $-23.10\pm0.31$& $1.55\pm0.20$ & $0.27\pm0.23$ &  1.82\\
&& F+NS & 20 & $-22.67\pm0.20$ & $1.29\pm0.15$ & $0.31\pm0.23$ &  1.89 \\
&& F+NS+S & 25 & $-22.31\pm0.16$& $1.07\pm0.13$ & $0.35\pm0.23$ &  2.34\\
\hline 
\multirow{3}{*}{$M(400)_{\rm 30}-M(520)_{\rm 30} \times \Delta M(400)_{\rm 30}$}  & \multirow{ 3}{*}{$M$(400)$_{\rm 0}$}& F & 12 & $-22.21\pm0.12$& $0.82\pm0.14$ & $0.19\pm0.19$ &  0.86 \\
&& F+NS & 14 & $-22.12\pm0.09$& $0.73\pm0.11$ & $0.18\pm0.16$ &  0.61\\
&& F+NS+S & 18 & $-21.95\pm0.08$& $0.60\pm0.11$ & $0.25\pm0.18$ &  1.03\\
\hline
\end{tabular}%
}
\end{center}
Least squares fits for a Bayesian weighted linear regression with weighted errors both in $x$ and $y$ of the form $\eta$ = $\beta$ + $\alpha\times x'$ + $\epsilon$, where  $x = x' + x_{\rm err}$ and  $y = \eta + y_{\rm err}$. The $\sigma$ is the standard deviation of this fit. The last column gives the reduced $\chi^2$.\\
\end{table*}

\section{SLSN I Standardisation}
We next confirm that the previous observed relationships between peak luminosity ($M(400)_0$) and decline rate in magnitudes over 30 days ($\Delta M(400)_{30}$), here referred to as peak decline, still hold. 
To do so, we first convert the rest-frame apparent magnitudes into absolute magnitudes (see Table~\ref{table:datarel}) using the same cosmology of previous studies (H$_0=72$ km/s, $\Omega_{\rm matter}=0.27$, $\Omega_{\Lambda}=0.73$) and employ a Bayesian approach to evaluate a weighted linear regression of these parameters, allowing for the uncertainties in both the $x$ and $y$ variables and intrinsic scatter \citep{ke07}. This process uses Bayesian inference that returns random draws from the posterior. Convergence to the posterior is performed using a Monte Carlo Markov Chain with $10^5$ iterations. 
A weighted regression provides a standard deviation bigger than the unweighted one by a factor of roughly $\sum_{i=1}^{n} 1/\sigma_{i}$, where $n$ is the sample size. For the peak decline relation and a sample of 20 objects we retrieve a standard deviation similar to that of the previous unweighted study \citep[$\sigma=0.33$,][]{in14}, suggesting that with a bigger sample we have decreased our scatter (see Table~\ref{table:fit}). This confirms that such relation is quantitatively useful to reduce the intrinsic scatter in the uncorrected peak magnitudes and hence provide a solid proof-of-concept that SLSNe~I may be used as cosmological standardisable candles.
The standard deviation of $\sigma=0.33$\,mag decreases, as expected, to $\sigma=0.26$\,mag if we use only the Fast subclass. Including the Slow subclass events substanially increases the dispersion to $\sigma=0.74$\,mag. Such a large dispersion further supports their exclusion.

We also retrieve a similar standard deviation to the previously published $M(400)_{\rm 0}$ vs. $\Delta (M$(400)$_{\rm 30}$ - $M$(520)$_{\rm 30}$) relation \citep{in14}. We also explore other possible correlations to check if an equally strong relation as those above mentioned can be found at a shorter time-scale (phase $<$ 30d). We do not find any strong correlation (see Table~\ref{table:fit}), but the $M(400)_{\rm 0}$ vs. $M$(400)$_{\rm 30}$ - $M$(520)$_{\rm 30}$  (peak - colour) relation provides the lowest $\widetilde{\chi}^2$  ($\widetilde{\chi}^2 = 0.90$) and $\sigma = 0.19$ mag for the F+NS sample. We also consider correlating both decline and colour information with luminosity ($M(400)_{0}$ vs. ($M(400)_{\rm 30}-M(520)_{\rm 30}) \times \Delta M(400)_{\rm 30}$), further reducing the scatter (see Table~\ref{table:fit}). However, the disadvantage of using such promising correlations is that they need a second, redder band (520\,nm). Hence the size and redshift coverage of the sample is smaller than that defined by the peak decline relation. Nevertheless, when the sample size becomes bigger than the current one ($\gtrsim$100), these two relations including the 520nm band might be more effective than the peak decline.

\section{Supernova model}\label{sec:cov}
We therefore use the peak decline standardisation method in our analysis, i.e. our distance estimator assumes that SLSNe~I with an identical light-curve decline rate have the same average intrinsic luminosity at all redshifts. The standardized distance modulus, $\mu_{\mathrm{obs}}$, is then given by
\begin{equation}\label{eq:std}
\mu_{\mathrm{obs}}= m(400) - M(400) + \gamma \Delta M(400)_{30}\,,
\end{equation}
where $m(400)$ is the peak apparent magnitude in rest-frame 400\,nm band, and $M(400)$ (the peak absolute magnitude) and $\gamma$ are nuisance parameters in the distance estimate. This is compared to the model distance modulus, $\mu_\mathrm{model}$, of $5\log_{10}(d_{\rm L}/10\,{\rm pc})$, where $d_L$ is the luminosity distance. The fit then minimizes the $\chi^2$ according to
\begin{equation}\label{eq:res}
\chi^2 = \Delta\vec{\mu}^T \cdot C^{-1} \cdot \Delta\vec{\mu}\,,
\end{equation}
where $C$ is the covariance matrix and $\Delta \vec{\mu}$ is the vector of residuals $\Delta \vec{\mu}=\mu_{\mathrm{obs}}-\mu_{\mathrm{model}}$. Note that the Hubble constant, $H_0$, enters in both $M(400)$ and $d_\mathrm{L}$, and thus does not affect (and is not constrained by) the cosmological fit. We assume an unperturbed Friedmann-Lema\^itre-Robertson-Walker metric and a flat universe, i.e., $\Omega_{\rm M} + \Omega_{\Lambda} = 1$, and the free parameters in the fit are therefore $\Omega_{\rm M}$ and the two nuisance parameters $M(400)$ and $\gamma$.
 
The covariance matrix is defined as the sum of the statistical and systematic parts
\begin{equation}\label{eq:cov}
C = C_{\rm stat} + C_{\rm sys}\,.
\end{equation}

The statistical covariance matrix is diagonal, while the systematic covariance matrix can contain off-diagonal terms capturing the covariance between different events. Here we define systematic uncertainties as those terms whose effect on our final uncertainty budget could not be reduced by increasing the size of the SLSN~I sample (i.e.  reddening, surveys zero points, Malmquist bias and the light curve fitting method). We note that, due to the limited size of our sample, our analysis is dominated by statistical uncertainties (i.e. uncertainties on fitted light curve parameters). 

In our minimization technique of Equation~\ref{eq:res} we use the following priors: the JLA result (SN analysis only) as a prior on $\Omega_{\rm M}$; the $M(400)$ and $\gamma$ values previously retrieved \citep{in14} for the $M(400)_{\rm 0}$ vs $\Delta M(400)_{\rm 30}$ as a prior of the other two nuisance parameters ($M(400)$, $\gamma$). We also assume a Gaussian distribution for the form of the priors. To minimize Equation~\ref{eq:res} we use {\sc iminuit}\footnote{\url{https://github.com/iminuit/iminuit}}, a minimization technique based on {\sc minuit} \citep{minuit} which runs over $10^5$ iterations, and a Markov Chain Monte Carlo Ensemble sampler \citep{emcee} with $5 \times 10^3$ iterations. Both provide non-Gaussian distributed uncertainties and similar results, although We note that {\sc iminuit} uncertainties are always a factor $\sim$1.5 bigger than those from {\sc emcee} and this is likely due to the small dataset. This is confirmed by the analysis executed with a bigger (847 objects) simulated dataset for which both algorithms give similar uncertainties (see Section~\ref{sec:future}).

\subsection{Statistical uncertainties}
The statistical uncertainties part is diagonal and includes uncertainties as follows:
\begin{equation}\label{eq:stat}
\begin{split}
C_{\rm stat,ii} & =  \,
\sigma^2_{m_{\rm 400},i} 
+ \gamma^2 \sigma^2_{\Delta M{\rm (400)}_{\rm 30}} \\
& 
+ \left( \frac{5\, (1+z_i)}{z_i\, (1+z_i/2)\, {\rm log}(10)} \right)^2 \times \sigma^2_{z,i} + \sigma^2_{\rm lensing}\,.
\end{split}
\end{equation}

Here $\sigma_{m_{\rm 400},i}$ and $\sigma_{\Delta M{\rm (400)}_{\rm 30}}$ are the uncertainties on the fitted light curve parameters. 
The 3rd term is associated with our choice of an empty-universe approximation for the relation between redshift uncertainty and the associated magnitude uncertainty \citep{davis11}. The last term is a random uncorrelated scatter due to lensing, $\sigma_{\rm lensing}=0.055 \times z $, following the prescription used for SNe Ia \citep{conley2011}. The lensing dispersion for point sources depends on the line of sight density distribution, not the source properties, so this is appropriate even though our SLSN~I population differs from the SNe Ia population. Further studies should address whether this functional form is appropriate for the high-redshifts in our sample, however even at $z=1.5$ the lensing dispersion is only $\sigma_{\rm lensing}\sim 0.08$ magnitudes.  Since this is an order of magnitude lower than the dispersion in magnitudes (the first two terms) and the mean lensing magnification should be zero, we consider any possible lensing bias to be negligible.

\subsection{Systematic uncertainties}
The definition of systematic uncertainties is not always unambiguous and it depends on labels given by authors \citep[e.g.][]{conley2011,betoule2014,scolnic2018}. Here we interpret them as those terms whose effects on our final uncertainty budget could not be reduced by increasing the SLSN~I sample. We also note that, due to the limited size of our sample, our analysis is dominated by statistical uncertainties and variations in the systematics have little leverage on our cosmological constraints. 

There is no standard method for handling supernova systematic effects, but the most common approach is to initially fit the dataset without any systematic effects (hence only with $C_{\rm stat,ii}$) and then marginalize over all the systematic terms by adding the systematic part of the covariance matrix to the statistical as in Equation~\ref{eq:cov}. The matrix is given by 
\begin{equation}
C_{\rm sys,ij} = 
\sigma^2_{{\rm reddening},i,j} + \sigma^2_{{\rm ZP},i,j} + \sigma^2_{{\rm MalmquistBias},i,j} 
+ \sigma^2_{\rm model,i,j}\, .
\end{equation}

\subsubsection{Reddening}
The first term is related to reddening and we account only for Milky Way reddening along the line of sight, including an estimated 10\% random uncertainty for each SLSN~I due to the conversion from dust column density to extinction \citep{schlegel98}. The extinction is always E($B-V)<0.02$ mag and for this low value the extinction law is almost insensitive to the choice of $R_V$, hence the assumption of a Galactic value of $R_V=3.1$ is appropriate \citep{cardelli89}. At this stage we do not consider host galaxy reddening (see discussion in Section~\ref{ss:red}) since SLSNe~I explode in dwarf galaxies with no reported host galaxy extinction for the majority of events \citep[$\sim$85\%, data from][]{lu14,le15,an16,pe16,ch17,schulze18}. When a host galaxy extinction value of E($B-V)>0.02$ is reported, that is due to spectral energy distribution (SED) modelling \citep{schulze18} rather than galaxy line analysis and hence exposed to larger uncertainties. 
However, in future analysis this term should be investigated more carefully and might be taken in consideration. 

\subsubsection{Zero points}
The second term is due to  the uncertainties in the zero points (ZP) of each survey in each filter and each field. For example, in the case of DES we have four different filters and ten SN deep fields, of which two are approximately one magnitude deeper in order to extend SN searches out to higher redshift.
However, the difference between the zero points of different fields is usually of the order of $10^{-3} - 10^{-4}$ mag \citep{desyr1}  and so are their general uncertainties. These uncertainties are at a mmag level and hence we consider a general uncertainty for each filter in each survey. The general ZP uncertainties are retrieved for DES \citep{desyr1} and PS1 \citep{tonry2012,schlafly12}. For the rest of literature SLSNe~I, which were not found by these surveys or did not have the majority of their data obtained from a single survey, we considered average\footnote{Average evaluated from the uncertainties of DES and PS1.} zero point uncertainties for each filter matching those reported by other large surveys \citep{desyr1,tonry2012,pessto}.

\subsubsection{Malmquist bias}
To model our search efficiency (i.e. how many SLSNe~I are missed), for each SLSN~I we simulated 10$^4$ light curves using a Monte Carlo approach. 
Due to the relatively small sample of our analysis we treat it as a simple smoothed offset between nearby ($z<0.4$), medium redshift ($0.4\leq z \leq 1.0$) and distant SNe ($z>1$). After correction for light curve shape, we find the magnitude offset to be 0.02 mag up to $z=1.0$ and less than 0.05 for the high-redshift objects of our sample. This is in agreement with the fact that the majority of the $z>1$ SLSNe~I were discovered before maximum light similar to those at lower redshift, suggesting that our overall sample is equally biased at all redshifts. However, a more in-depth treatment of the Malmquist bias might be needed with bigger samples. That should make use of the predicted number of SLSNe~I for an unbiased survey to the number observed as a function of redshift, for which DES is an ideal test-bed \citep[][Thomas et al. in prep.]{angus18}. 

\subsubsection{Light curve fitting}
The last term, $\sigma_{\rm model}$, relates to our uncertainties in the interpolation used to fit our light curve, which means the kernel chosen for the GPs. We use a Matern-3/2 kernel that we find a suitable choice to avoid over-fitting and retrieve a balanced precision/recall outcome \citep{in18c}. However, other kernels can be used, such as the Matern-5/2, and we have to take into account the differences in the light curve outputs with different kernels. This term, in principle, could be reduced with more SLSNe~I, but is correlated between different SLSNe~I and therefore cannot be included in $C_{\rm stat,ii}$. 

\begin{figure*}
\centering
\includegraphics[width=16cm]{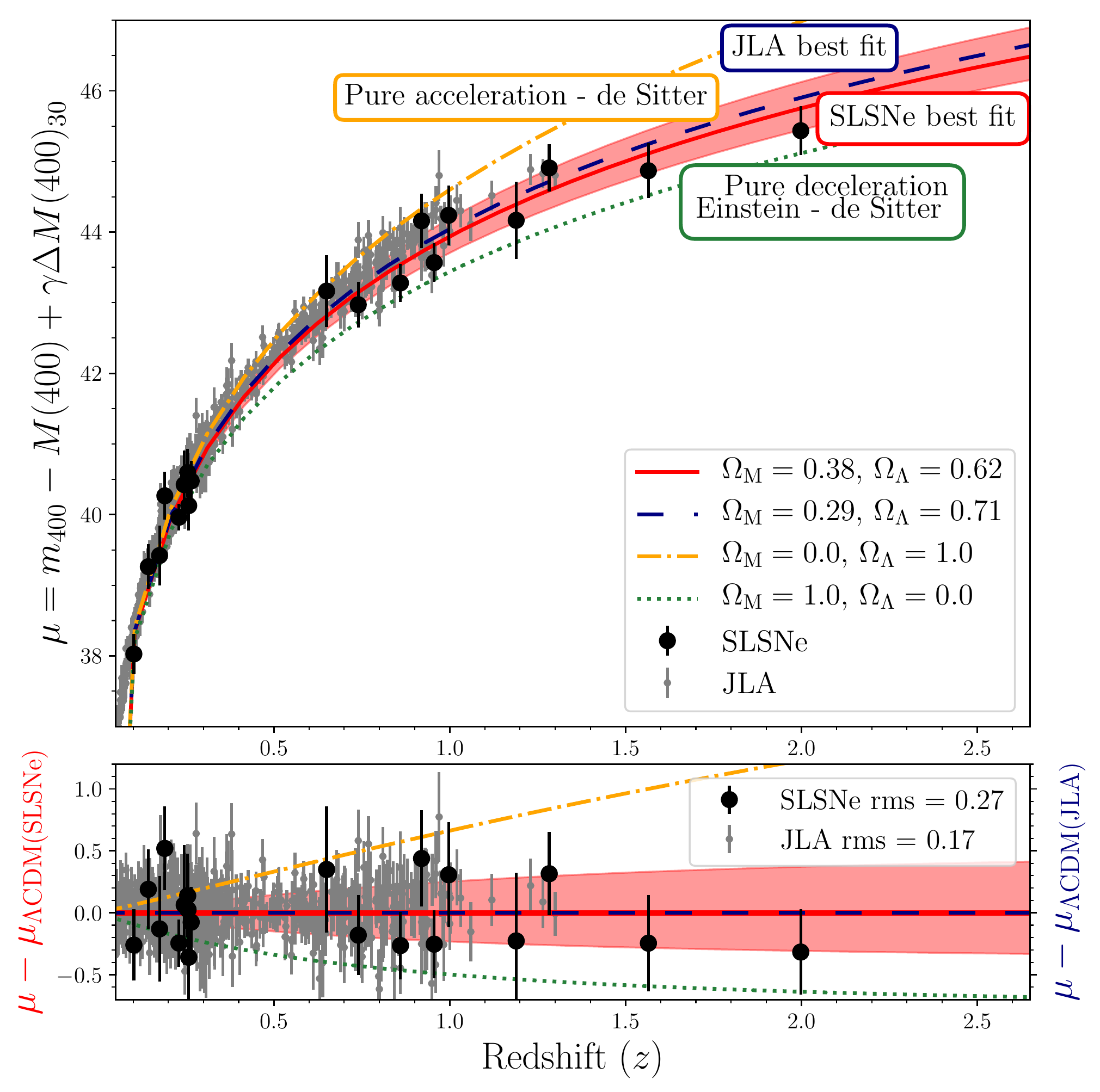}
\caption{Upper panel: The Hubble diagram of our SLSN~I sample (only F+NS subtypes) using the $\Delta M$(400)$_{\rm 30}$ standardization method. Over-plotted (solid red line) is the best fitting flat $\Lambda$CDM cosmology and its uncertainties (shaded red area) with $\Omega_{\rm M}=0.38^{+0.24}_{-0.19}$, measured only using SLSNe~I. Lower panel: the residuals of each SLSN~I from the best-fit cosmology (red line w.r.t. red left label) as a function of redshift. The JLA SN Ia compilation from their best-fit (dashed blue line) and residuals (grey dots vs blue line) are also shown as comparison. We chose the JLA sample because both studies use the same approach to derive the cosmological constraints, unlike the most recent Pantheon sample.}
\label{fig:hubble}
\end{figure*}

\subsubsection{Additional uncertainties}
We do not introduce any additional term for contamination from other SN types since all our objects have spectroscopic confirmation and have passed the 4OPS criterion. We also did not include any systematic related to peculiar velocities or a discontinuous step in the local expansion \citep[Hubble bubble,][]{jha07} since all our SLSNe~I have $z>0.1$. We also do not include any correction from host-galaxy properties since they all reside in similar galaxies at both low and high redshift \citep{le15} and no mass step function has been currently observed for SLSNe~I as has been seen for type Ia \citep{sullivan10}. 
A possible differential evolution with galaxy mass at high redshift was recently claimed in an analysis using rest-frame optical data of SLSN host galaxies \citep{schulze18}. However, this analysis does not hold if the rest-frame is extended to include wavelengths bluer than the $B$-band. 
Hence, we do not consider any additional source of uncertainties linked to galaxy evolution.

\begin{table*}
\caption{Best-fit parameters for the flat $\Lambda$CDM model using SLSNe~I alone and using the $M$(400)$_{\rm 0}$ vs. $\Delta M$(400)$_{\rm 30}$ standardisation relation.\label{table:hfit}}
\begin{center}
\resizebox{17cm}{!}{%
\begin{tabular}{lccccccc}
\hline
\hline
Sample& $\Omega_{\rm M}$ &$M$ (value) & $\gamma$  &Residuals (rms)  &$\Lambda=0$ Residuals (rms) &Redshift& $\chi^2$/d.o.f. \\
\hline
F+NS (stat+sys)& $0.38^{+0.24}_{-0.19}$&  $-22.40^{+0.38}_{-0.35}$ & $0.62^{+0.19}_{-0.19}$  & 0.27 &0.31&$\sim2.0$ &11.1/18\\ 
F+NS (stat)& $0.37^{+0.31}_{-0.19}$& $-22.45^{+0.38}_{-0.35}$ & $0.63^{+0.19}_{-0.19}$   &  0.25&0.27& $\sim2.0$&11.0/18 \\ 
F (stat+sys)& $0.22^{+0.29}_{-0.21}$&  $-23.20^{+0.56}_{-0.53}$ & $0.97^{+0.25}_{-0.25}$  & 0.26& 0.40& $\sim1.2$&6.4/13\\ 
F (stat)& $0.41^{+0.22}_{-0.20}$&  $-22.83^{+0.43}_{-0.41}$ & $0.88^{+0.22}_{-0.22}$  & 0.21 &0.26& $\sim1.2$&4.9/13\\ 
\hline
\end{tabular}%
}
\end{center}
\end{table*}

\section{SLSNe~I Hubble diagram}\label{sec:hubble}
Our sample size (20 F+NS SLSNe) is sufficient to provide a constraint on a single parameter driving the evolution of the expansion rate. In particular, in a flat universe with a cosmological constant (hereafter flat $\Lambda$CDM), SLSNe~I alone can provide a measurement of the reduced matter density $\Omega_{\rm M}$. The SLSN~I Hubble diagram and the flat $\Lambda$CDM best fit, derived from the minimization above, are shown in Figure~\ref{fig:hubble}. The fit parameters are given in the first row of Table~\ref{table:hfit}. We find a best fit value of $\Omega_{\rm M}=0.38^{+0.24}_{-0.19}$ and $\mathrm{rms}=0.27$\,mag for the residuals of the distance moduli, also shown in Figure~\ref{fig:hubble}. For comparison, the dispersion in the Joint Light curve Analysis (JLA) SN Ia sample is $\mathrm{rms} = 0.17$\,mag for 740 SNe Ia over $0.01<z<1.2$ \citep{betoule2014}.  This is also shown in Figure~\ref{fig:hubble}, where the JLA sample and its residuals are overplotted. The redshift coverage makes it possible to assess the overall consistency of the SLSN~I data with the flat $\Lambda$CDM model. 

\begin{figure}
\center
\includegraphics[width=\columnwidth]{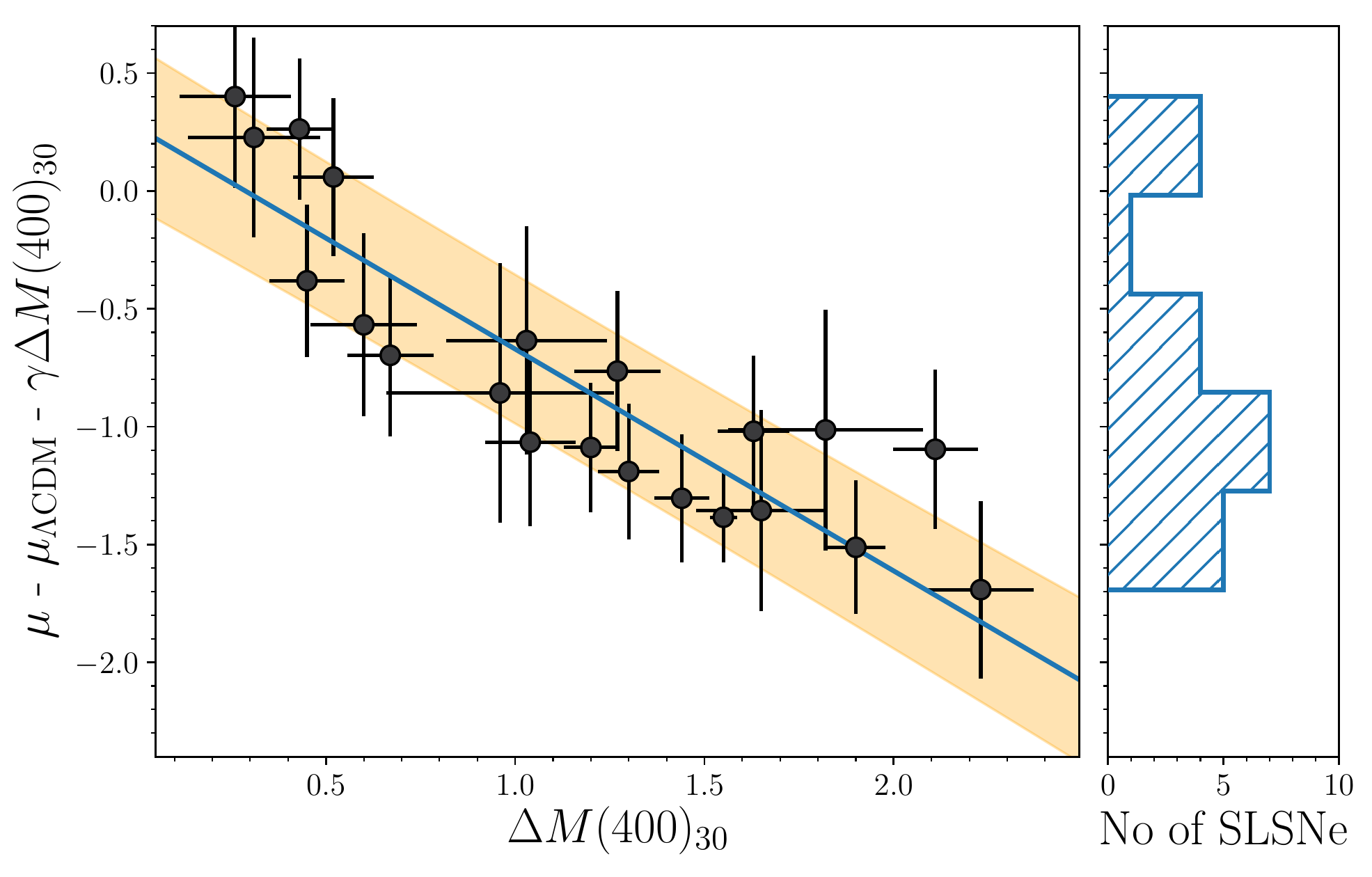}
\caption{Left: The SLSN~I residuals uncorrected for the decline parameter, plotted as a function of $\Delta M$(400)$_{\rm 30}$. This diagram computes distance modulus and returns the brighter-slower relationship. Right: histogram of the data distribution. The bin dimension has been chosen according to the Freedman Diaconis Estimator, which accounts for data variability and data size, and is optimised for smaller datasets.}
\label{fig:relnocosmo}
\end{figure}

In Figure~\ref{fig:relnocosmo} we plot the residuals of our sample without the decline-rate correction, as a function of the decline rate ($\Delta M$(400)$_{\rm 30}$). This shows the brighter-slower relationship of the standardization, with no apparent evolution in residuals across this relationship for our SLSN~I sample. We search for any further significant trends between decline, colour, and Hubble residuals, but find none. Thus, if further parameters are capable of decreasing the scatter in the residuals of our cosmological fit, they are either related to quantities that we have not measured, or larger samples are required to investigate them.

\section{$w_0 w_a$CDM cosmology with SLSNe I}\label{sec:ww}
We also explore how and to what extent this SLSN~I sample can improve the constraints on the redshift dependent equation-of-state of dark energy \citep{descosmo}, $w(a) = w_0 + (1-a)w_a$, where $a=(1+z)^{-1}$ is the cosmological scale factor. In our analysis, we include priors on the cosmological parameters from measurements of the CMB from \textit{Planck} \citep{planck16}.
We assume a Gaussian distribution for the form of the prior, which we construct at the maximum likelihood value used by the \textit{Planck} consortium \citep{planck16}. We also include the JLA SNe Ia sample \citep{betoule2014}. We adopt a flat $w_0w_a$CDM cosmological model, and a SLSN~I likelihood of the form $\ln \mathcal{L}\propto\chi^2$, where the $\chi^2$ is given by Equation~\ref{eq:res}, and $\mu_{\mathrm{model}}$ is now a function of $w_o$, $w_a$, $\Omega_{\rm M}$, and H$_0$.  To obtain convergence we use the
CosmoMC tool \citep{cosmomc,cosmomc2}

\begin{figure}
\centering
\includegraphics[width=\columnwidth]{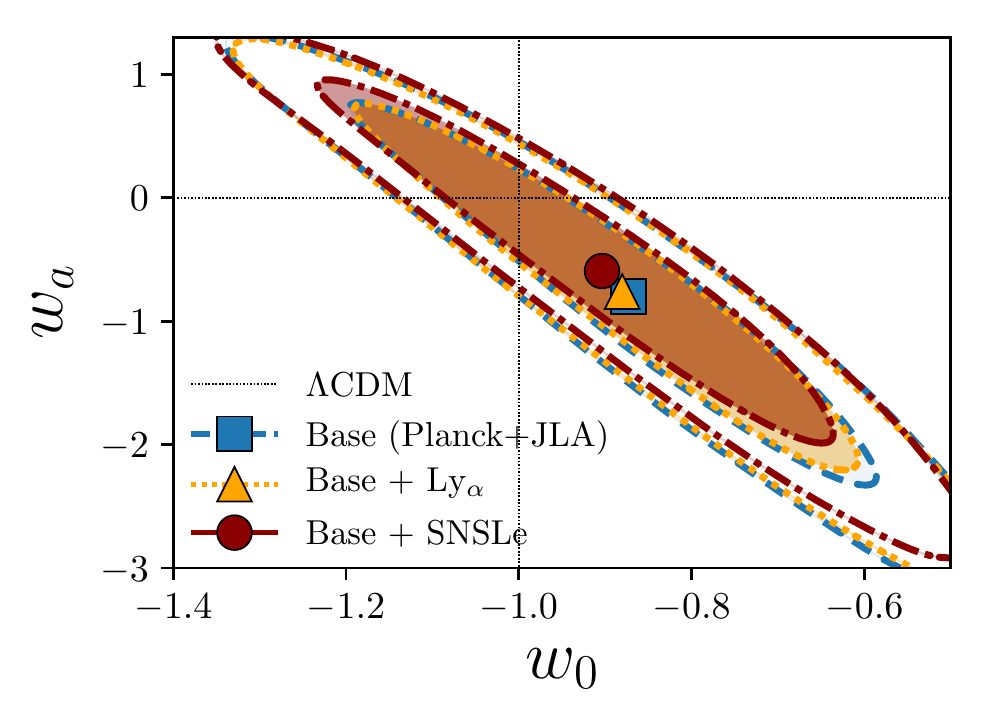}
\caption{Constraints on the dark energy equation-of-state parameters $w_0$ and $w_a$. We illustrate the one (filled) and two (unfilled) standard-deviation contours, and the maximum likelihood values for the `Base' configuration (JLA+CMB; blue dashed lines and square centroid) for estimates calculated including SLSNe~I (dark red dot-dashed lines and circle centroid), or Lyman-$\alpha$ BAO (orange dotted lines and triangle centroid). The horizontal and vertical dashed lines  at $w_0=-1$ and $w_a=0$ correspond to a cosmological constant. The Base+SLSNe~I  results are $w_0 = -0.904^{\pm 0.185}$ and $w_a = -0.594^{\pm 0.926}$, which is a small improvement of 4\% with respect to the joint CMB+JLA dataset, similar to that obtainable with the Base+Lyman-$\alpha$ configuration.}
\label{fig:w}
\end{figure}

We find marginalised constraints on $w_0 = -0.904^{\pm 0.185}$ and $w_a = -0.594^{\pm 0.926}$, which are shown in Figure~\ref{fig:w}.
To measure the improvement on the constraints on $w_0$ and $w_a$, we evaluate the Figure of Merit (FoM) proposed by the Dark Energy Task Force \citep{2006astro.ph..9591A}, which is the area enclosed by the two-standard deviation contour in the $w_0$-$w_a$ plane ($1/(\sigma_{w_0} \times \sigma_{w_a})$). Without SLSNe~I, the FoM is 5.622, which increases by 4\% to 5.835 with the addition of SLSNe~I. We then compare this improvement with constraints from Lyman-$\alpha$ forest baryonic acoustic oscillations \citep{lyalpha}, which is also a high-redshift probe like SLSNe~I. The FoM for the CMB+JLA+Lyman-$\alpha$ is 5.9, suggesting that at present SLSNe are comparable to Lyman-$\alpha$ BAO at such redshifts. That is also true of we analyse the uncertainties of the three data combinations (Base, Base+SLSNe, Base+Lyman-$\alpha$) on $w_0$ and $w_a$ which are 0.187, 0.185 and 0.183 ($\sigma_{w_0}$) and 0.952, 0.926, 0.928 ($\sigma_{w_a}$). 

\begin{figure*}
\center
\includegraphics[width=16cm]{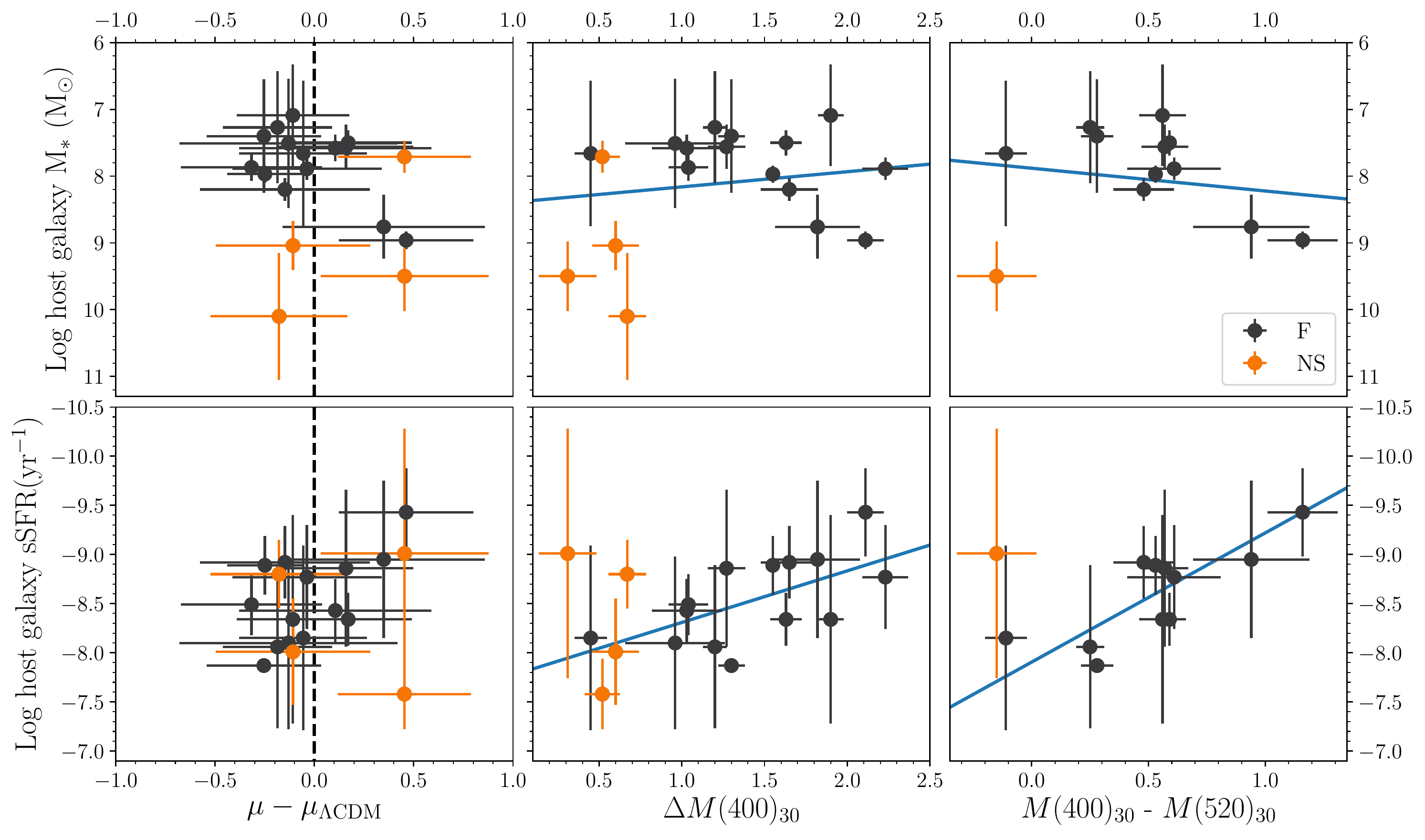}
\caption{SLSNe~I residuals from the best-fit flat $\Lambda$CDM cosmology (left panels), decline over 30 days (middle  panels) and colour at 30 days (right panels) as a function of host galaxy M$_*$ (upper row) and sSFR (bottom row). The error bars on the individual SLSNe~I are taken from the SED fitting for the sSFR and M$_*$ axes and are the statistical errors propagated through the light curve fitting for the residual and observables axes. Bayesian weighted linear regression (blue solid line) are also displayed with the exception of the residuals.}
\label{fig:environment}
\end{figure*}

\section{Environment}\label{sec:enviro}
We investigate if there is an additional dependence on the global
characteristics of SLSN~I host galaxies. We retrieve SLSN~I host galaxies information from \citet{schulze18}, namely specific star formation rate (sSFR) and stellar mass (M$_{*}$), which used the SED fitting algorithm {\sc lephare} \citep{arnouts99,ilbert06} and a Chabrier initial mass function \citep{chabrier03}. This approach gives almost identical results to the {\sc magphys} SED fitting \citep{magphys} as shown by the SLSN~I dataset in literature \citep{ch17,schulze18}. We then apply the same approach to the only DES SLSN~I with host galaxy photometry that was not presented in the \citet{schulze18} sample \citep[DES16C2nm,][]{mat18}. 
However, we note that the broad-band SED fitting approach used here is a relatively crude way to determine galaxy properties and hence this analysis is only an initial investigation into host galaxy dependencies. 

In Figure~\ref{fig:environment} (left panels) we compare host galaxy properties with our residuals, but we do not find a mass step function (or any relation). We observe mild correlations between the host galaxy sSFR and the light curve properties in terms of decline (Pearson {\it r} = 0.45) and colour (Pearson {\it r} = 0.52). This suggests that SLSN~I in low-sSFR host galaxies are redder and faster decliners than those in high-sSFR. There is no appreciable trend with the stellar mass of host galaxies. From this analysis it seems that there is not a systematic uncertainty in the residuals introduced by SLSN~I host properties, but further analysis with a bigger sample and a more precise estimates of global host properties and local to the SN environment is encouraged. 

\begin{figure*}
\center
\includegraphics[width=16cm]{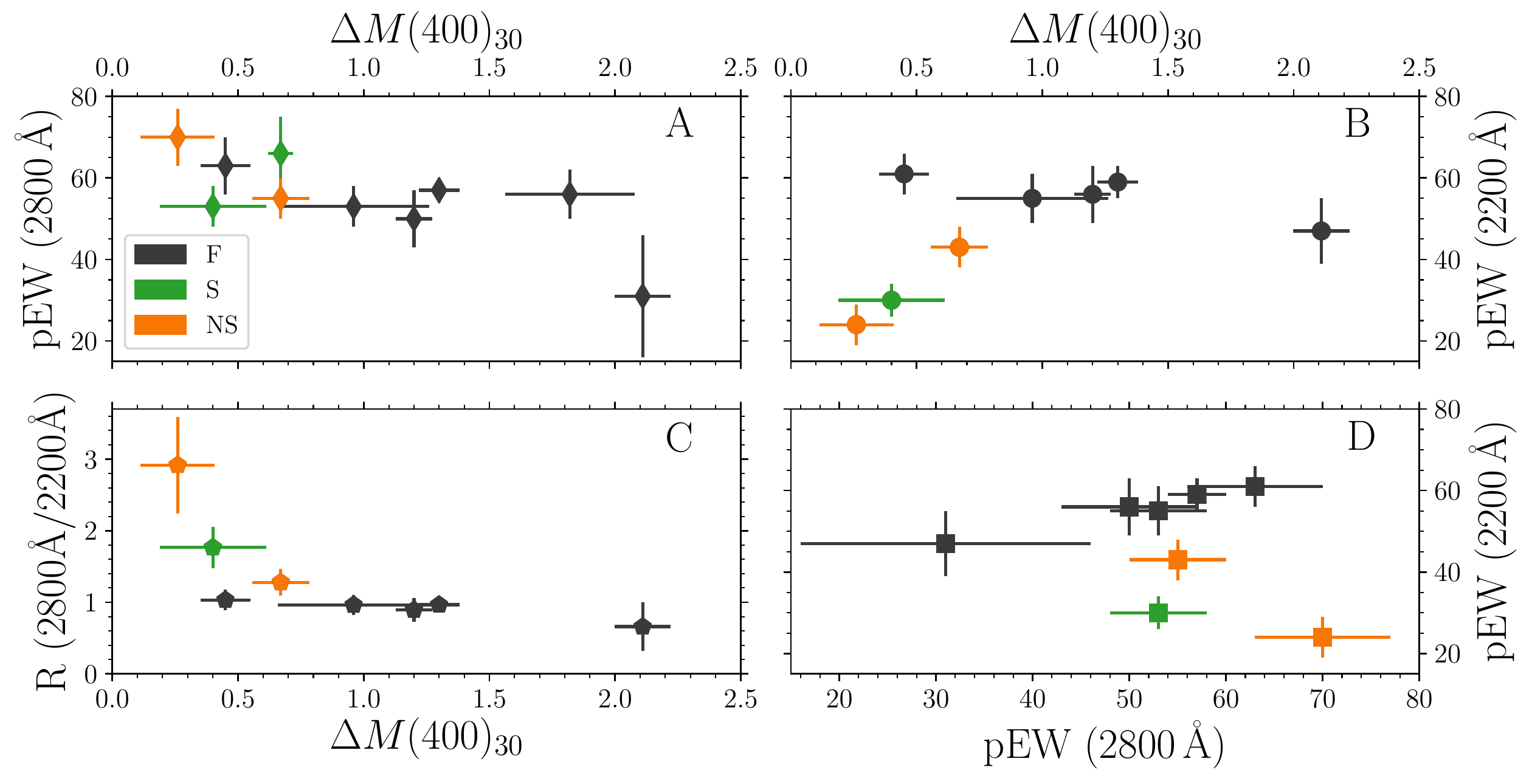}
\caption{Spectroscopic, pEW(Mg~{\sc ii}/C~{\sc ii}) at 2800~\AA\/, pEW(C~{\sc iii}/C~{\sc ii}/Ti~{\sc iii}) at 2200~\AA\/ and their ratio (R) versus photometric, $\Delta M$(400)$_{\rm 30}$, measurements. A mild trend is displayed in panel A (diamond markers), with a Pearson {\it r} coefficient of 0.76. There are promising relationships involving pEW(C~{\sc iii}/C~{\sc ii}/Ti~{\sc iii}) in panels B and D, although a larger UV sample is required to confirm this.}
\label{fig:uv}
\end{figure*}

\section{Exploring SLSNe I in the ultraviolet}\label{sec:uv}
Motivated by studies exploring the velocity and shape of lines of ionised elements in the UV \citep[e.g.][]{galyam19}, we measure the pseudo equivalent width (pEW) of the C~{\sc iii}/C~{\sc ii}/Ti~{\sc iii} and Mg~{\sc ii}/C~{\sc ii} blended lines at $\sim2200$ \AA\/ and $\sim2800$ \AA\/ respectively, to look for a more quantitative method of distinguishing between Fast and Slow subtypes at high redshift. We check if, combining the UV line pEWs with light curve evolution around peak (-10 d $<$ phase $<$ +30 d), we can find similar clusters  to those observed in the optical, using Fe~{\sc ii} lines. The prefix ``pseudo'' is used because the reference continuum level chosen does not represent the true underlying continuum level of the supernova. It defines the strength of the line with respect to the pseudo-continuum at any given time. 
We choose the  Mg~{\sc ii}/C~{\sc ii} line since it is easy to sample at $0.1\lesssim z \lesssim 3.0$, which covers this dataset and the majority of future datasets (see Section~\ref{sec:future}, and it is a good proxy for the outermost layers of the carbon/oxygen-rich material \citep{maz16}.  The C~{\sc iii}/C~{\sc ii}/Ti~{\sc iii} line is also a good proxy, when available, and it is usually stronger than Mg~{\sc ii}/C~{\sc ii}. Such difference in strengths is likely due to a lower excitation potential of the lines at 4200~\AA\/ and a bigger contribution to the blending from carbon lines. Nevertheless, it is  harder to sample for our redshift baseline.

We collected 10 SLSNe~I sampling these UV lines \citep{barbary09,cho11,mm14,vr14,lu16,yan17,mat18,quimby18,angus18}. Eight of them also show the Fe~{\sc ii} $\lambda 5169$; six were identified as Fast, and two as Slow. The other two do not have Fe~{\sc ii} lines sampled due to their higher redshift and are labelled as NS. For six of them (4F+2S) we have both pre- and post peak spectra. We measure the pEW \citep[see][for further details in the methodology]{claudia17a} and for each SN we do not observe any change in the pEW values from $-10$ days to $+10$ days. Hence we group our measurements as `at peak epoch' ($-10$\,d~$<$~phase~$<$~$+10$\,d). We also measure the line velocities and find $18000 < v\,{\rm (km/s)} < 23000$ in agreement with previously results \citep{cho11,vr14,maz16}. However, due to the blending of several ions, we decide to only focus on pEWs, which are less sensitive to the signal-to-noise and flux-calibration issues than velocities, flux ratios and line depths \citep{folatelli13}.

We then compare the pEWs and their ratio (R = pEW(Mg~{\sc ii}/C~{\sc ii})~/~pEW(C~{\sc iii}/C~{\sc ii}/Ti~{\sc iii})) with the light curve decline ($\Delta M$(400)$_{\rm 30}$). 
In Figure~\ref{fig:uv} no clear groups are observed comparing the pEW(Mg~{\sc ii}/C~{\sc ii}) or the ratio with the decline, panel A and C respectively. Instead, a mild correlation (Pearson {\it r}~$=0.76$) is shown in panel A. When comparing the pEW(C~{\sc iii}/C~{\sc ii}/Ti~{\sc iii}) with the decline (panel B) or with the pEW(Mg~{\sc ii}/C~{\sc ii}) in panel D, we retrieve promising clusters. However, the dataset is rather small and no further analysis can be done, although in future the pEW(Mg~{\sc ii}/C~{\sc ii}) versus pEW(C~{\sc iii}/C~{\sc ii}/Ti~{\sc iii}) analysis could give useful information for their characterization at high redshift ($z\gtrsim0.8$). 

In the future we might use such UV information to add another nuisance parameter, e.g. pEW(Mg~{\sc ii}/C~{\sc ii}), to Equation~\ref{eq:std}. This additional parameter would transform Equation~\ref{eq:std} into 
\begin{equation}
\begin{split}
\mu &= m_{400} - M(400) + \alpha \\
& \times (\Delta M(400)_{\rm 30} + \beta \times {\rm pEW(Mg~II/C~II))}\,,
\end{split}
\end{equation}
with three nuisance parameters ($\alpha$, $\beta$, $M(400)$) and would allow SLSNe~I Slow to be included in the standardization. Although this approach is appealing, we need a larger UV spectroscopic dataset to confirm the findings described above.

\begin{figure}
\center
\includegraphics[width=\columnwidth]{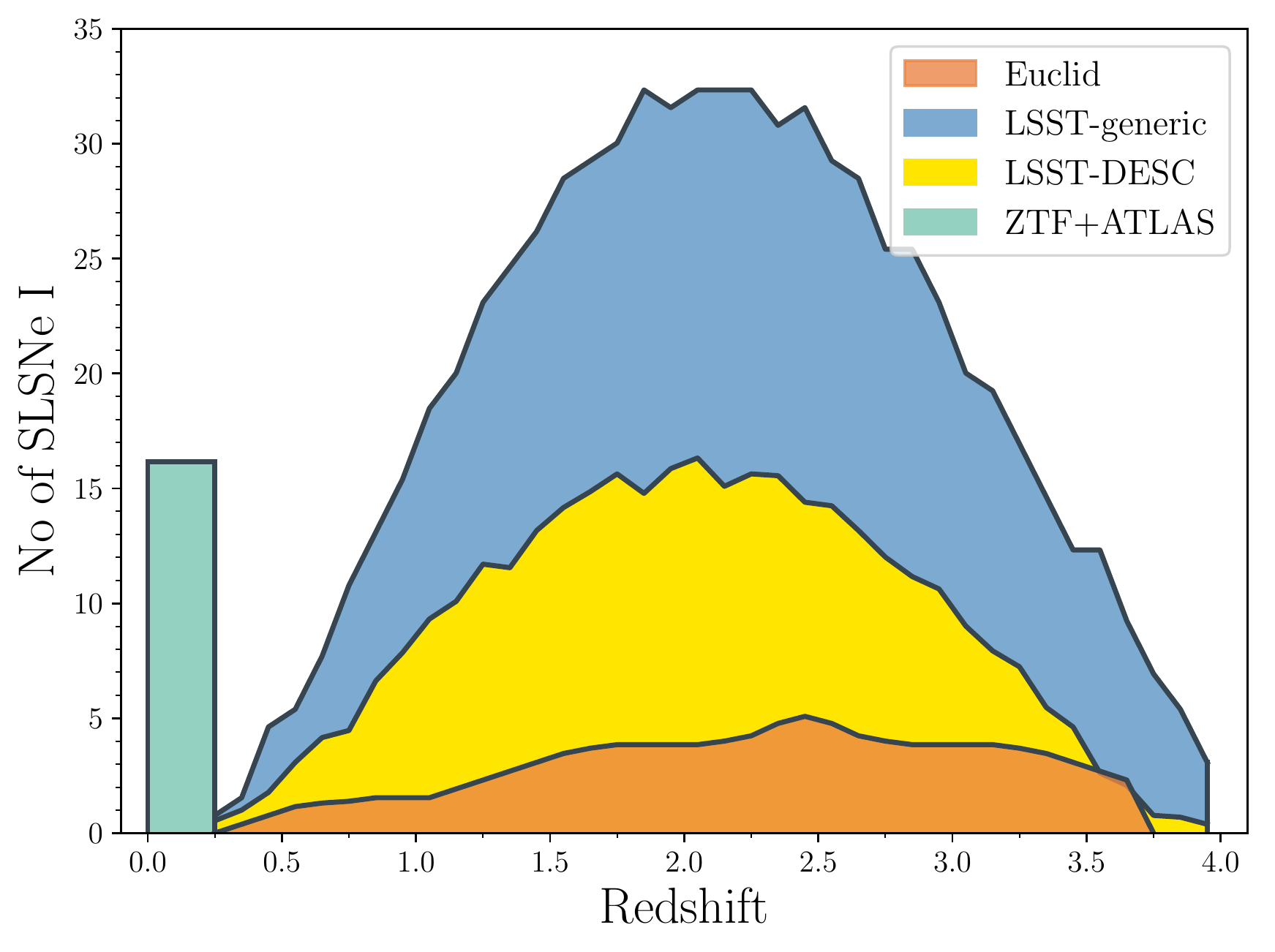}
\caption{Predicted distribution of SLSNe~I as a function of redshift. Bins are $\Delta z =0.1$ for the LSST deep drilling fields SLSNe, while the Euclid rates, binned with a $\Delta z =0.5$ \citep{in18a}, have been here re-sampled with a $\Delta z =0.1$. A flat SLSN~I distribution up to $z\sim0.3$ from ZTF+ATLAS with a $\Delta z =0.1$ bin size is also shown (see Section~\ref{sec:future}).}
\label{fig:futdistr}
\end{figure}

\begin{table*}
\caption{SLSN~I future simulated dataset out to redshift $z=3.5$, from two configurations of the LSST deep drilling fields, the {\it Euclid} satellite, and the low-redshift ZTF and ATLAS surveys. Results for the flat $\Lambda$CDM model (i.e., $\Omega_{\rm M}$), and the $w_0w_a$CDM model Figure of Merit are also reported\label{table:future}. Precision on $\Omega_{\rm M}$, due to statistical uncertainties only, has been increased as a consequence of adopting a larger sample.}
\begin{center}
\begin{tabular}{lcc}
\hline
\hline
Surveys & No. of SLSNe~I & $\Omega_{\rm M}$ \\
\hline
ZTF + ATLAS + LSST-generic + {\it Euclid}  &  868 & $0.278^{+0.018}_{-0.017}$\\
ZTF + ATLAS + LSST-DESC + {\it Euclid}  &  492 & $0.269^{+0.022}_{-0.021}$\\
\hline
\hline
Configuration & \multicolumn{2}{c}{Figure of Merit (FoM)}\\
\hline
CMB+JLA &  \multicolumn{2}{c}{5.60}\\
CMB+JLA+Lyman-$\alpha$ &  \multicolumn{2}{c}{5.90}\\
CMB+JLA+SLSNe &  \multicolumn{2}{c}{5.84}\\
CMB+JLA+SLSNe(ZTF + ATLAS + LSST-generic + {\it Euclid}) &  \multicolumn{2}{c}{15.50}\\
CMB+JLA+SLSNe(ZTF + ATLAS + LSST-DESC + {\it Euclid}) &  \multicolumn{2}{c}{11.50}\\
\hline
\end{tabular}
\end{center}
\end{table*}

\section{SLSNe I cosmology: future and improvements}\label{sec:future}
To understand the future potential of SLSNe~I in cosmology, we consider SLSN~I rates for the {\it Euclid} satellite \citep[135 high-quality SNe in 5 years,][]{in18a}, and SLSN~I predictions for the deep drilling fields of the Legacy Survey of Space and Time (LSST) at the Vera Rubin Observatory. At the moment of writing this paper, the final cadence and number of deep drilling fields are still under evaluation. Hence, we assume two configurations. A generic set-up with 10 deep drilling fields, visited 180 days each year with a 5 day {\it griz} cadence. The single visit depths are 25.0, 24.7, 24.0, 23.3 mag in {\it griz}, respectively \citep[AB magnitudes for a $5\sigma$ point source,][]{lsst}. The second configuration is that proposed in the white paper of the Dark Energy Science Consortium (DESC) at the Vera Rubin Observatory. This proposes 5 deep drilling fields, multiple visits per night in {\it griz} a 4 day cadence and depths of 25.3, 25.0, 24.8, 24.5 mag.
Following the methodology of the previous work of \citet{pr17}, we use an average peak SLSN I luminosity of $M(400)_0 = -21.756 \pm 0.495$ mag (where the average and standard deviations are determined from combing the F+NS subgroups), a model spectral energy distribution and a spectral template for $k$-correction. 
This method is consistent with that previously used  to predict the number of SLSN~I in the LSST wide survey \citep{sco16}. Here, we only consider SLSNe~I that have been detected four times in at least three filters which is, for consistency, the same that has been done for the {\it Euclid} SLSN~I rates \citep{in18a}. We then retrieve 929 and 441 SLSNe~I in the range $0.25<z<3.95$ for the `generic' and `DESC' configuration, respectively. We note that even with an unfavourable LSST cadence to discover and monitor transients, such as normal supernovae, we would expect to recover SLSN~I at $z>1$ due to their intrinsic high-luminosity and slow-evolution which will be further exaggerated due to time dilation in the observer frame \citep{in18a,moriya18}.  We note that our results are not as optimistic as those published in \citet{villar2018}. However, the methodology followed here is based on an observed luminosity distribution while that of \citet{villar2018} is based on a prescription of a magnetar model which has been shown to provide non-physical results or discordant values with those of other prescriptions \citep[see Table A1 in][]{ni17b}. 

We also make predictions on the number of suitable SLSNe~I that will be observed by the Zwicky Transient Factory \citep[ZTF,][]{bellm14} and the Asteroid Terrestrial-impact Last Alert System \citep[ATLAS,][]{atlas} at $z<0.25$.
We assume the low-redshift SLSN~I rates reported in literature \citep[][which include Slow events]{qu13,pr17} and scale the star formation history \citep{li08} accordingly to construct a volumetric rate evolution. Based on assumptions of the ZTF and ATLAS observing strategy \citep{bellm14,smith2020}, and a 5${\sigma}$ depth of 20.5\footnote{ZTF would need a depth of 22.66 mag to catch all SLSNe~I out to z=0.5.} and 19.5 mag respectively, we calculate the number of events ZTF and ATLAS would be capable of following to measure a decline of $\sim2$ mag from peak. We retrieve a very conservative number of 21 SLSNe~I per year out to $z\sim0.25$\footnote{From the ZTF/ATLAS ATels and TNS AstroNotes we confirm that this number is conservative.}, with information on the decline. Both surveys are expected to run for at least 3 years, and hence we expect a total of at least 63 SLSNe~I from them.

Considering the observed number of SLSNe~I \citep{in19} and the relative fractions of Fast and Slow subgroups determined from the statistical analysis of the SLSN I population, in our simulated dataset we envisage a division of 58\% Fast, 23\% Slow and 19\% with No Subclass. Hence our simulated samples have a total of 868 and 492 useful SLSNe~I F+NS for our analysis.
We run Monte Carlo simulations with 868 and 492 SLSNe~I (LSST+{\it Euclid}+ZTF+ATLAS, see Table~\ref{table:future}) following the redshift distribution of Figure~\ref{fig:futdistr} ($0.02<z<3.95$). We randomly place them into the relation of Figure~\ref{fig:relnocosmo} within 3$\sigma$ from the best-fit ($x_i,y_i$), and not within the 2.2$\sigma$ discussed above (see Section~\ref{sec:sample}), to account for increased uncertainties in the identification of SLSN~I subclasses at high redshift. We associate random uncertainties in both $x$ and $y$ ($x^{\rm err}_{\rm min}<x^{\rm err}_i<x^{\rm err}_{\rm max}$ and $y^{\rm err}_{\rm min}<y^{\rm err}_i<y^{\rm err}_{\rm max}$). We also assign a random distance for each redshift bin. Using this set-up, we run our cosmological fitter as previously done (see Section~\ref{sec:hubble}) and retrieve $\Omega_{\sc M}=0.262^{+0.020}_{-0.018}$, where the uncertainties are only statistical.
With a similar size dataset to those of current type Ia cosmology (e.g. Pantheon and JLA), we also retrieved statistical uncertainties of the same order of those achieved using SNe Ia. This is promising since at our current stage it seems that statistical uncertainties are the major contributor to the SLSN~I cosmology error budget.
Estimating systematic sources of uncertainty which might occur at high redshift, such as dust evolution, is beyond the scope of this study but the occurrence of SLSNe~I almost exclusively in low metallicity environments might suggest of a typically low dust content for the vast majority of these events \citep{wiseman17}. 
With this set-up we also explored the $w_0 w_a$CDM cosmology and found that the Figure of Merit of CMB+JLA+SLSNe(ZTF + ATLAS + LSST-DESC + {\it Euclid}) and CMB+JLA+SLSNe(ZTF + ATLAS + LSST-generic + {\it Euclid})are 11.5 and 15.5, respecitvely. While that of CMB+JLA is 5.6 suggesting that in the future, SLSNe~I will help deliver more precise cosmological constraints (of a factor of 2-3) than this proof-of-concept.
This analysis suggests that with a sample almost as large as that of current type Ia cosmology \citep[e.g. the Patheon sample,][]{scolnic2018}, we can retrieve a similar statistical precision for $\Omega_{\rm M}$ and can independently confirm type Ia findings, as well as reach a redshift range that should be matter-dominated but still unexplored with type Ia SN cosmology.

\begin{figure}
\centering
\includegraphics[width=\columnwidth]{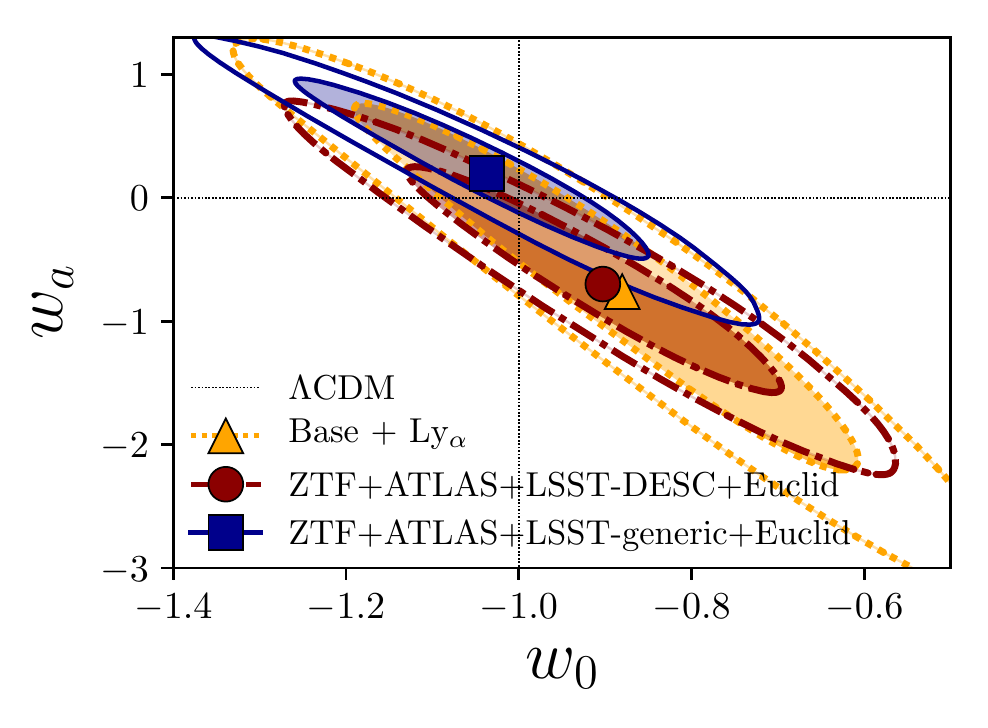}
\caption{Constraints on the dark energy equation-of-state parameters $w_0$ and $w_a$ as for Figure~\ref{fig:w} where the Base configuration is given by CMB+JLA. We compare the Base+Lyman-$\alpha$ configuration with the two future BASE+SLSNe configurations which both show the improvement in precision of the cosmological analysis (see the FoMs in table~\ref{table:future}).}
\label{fig:wf}
\end{figure}

A possible issue might be selecting 868 (or 492) cosmological useful SLSNe~I among the total number of SLSNe~I since optical spectroscopy can probe only Fe~{\sc ii} line out to $z\sim1$ and hence identifying their subtype would be challenging. In future, this may be solved with the European Extremely Large Telescope (E-ELT) and the High Angular Resolution Monolithic Optical and Near-infrared Integral field (HARMONI) spectrograph (first light in 2025) which is capable to probe the rest-frame region of interest out to $z\sim3.7$ \citep{hsim}. However, it has recently been shown that UV absorption lines in SLSNe~I Fast are generally sharper than Slow, in contrast to what happens to the absorption lines of less ionised elements in the optical. However, in SLSNe~I Fast, all species at $\lambda > 1500$\AA\/ show faster velocities than in Slow events \citep{galyam19}. Such behaviour can also be appreciated in the spectroscopic UV comparison between low redshift SLSNe~I and the high redshift SLSN~I at $z\sim2$ \citep{mat18}, where it has been demonstrated that SLSNe~I Fast recede faster in the UV than Slow events, as well as displaying an irregular UV colour evolution.
Such results have not yet been confirmed with a larger dataset due to the paucity of UV observations. However, they are somewhat reassuring, as they may be used to pave the way to the identification of cosmological useful SLSNe~I by means of optical facilities up to the highest redshift of our predictions. Thanks to their characteristic light curve evolution \citep[][Inserra \& Parrag in prep.]{mat18,in19} and more accurate machine learning techniques \citep[e.g.][]{moller19,ishida19,mutu19}, SLSN I identification might be possible without the need of spectroscopy. We note that considering and measuring an uncertainty parameter for high-redshift, photometrically identified SLSNe~I is premature and beyond the scope of this work.

\section{Conclusions}
We examined a sample of 26 SLSNe~I, 20 of which are useful for a cosmological analysis. We confirmed the previously established standardization relation of SLSNe~I \citep{in14} with a larger dataset and improved light curve fitting technique, and used the sample to make a measurement of the cosmological parameter $\Omega_{\rm M}$. The resulting Hubble diagram contains the highest spectroscopically-confirmed redshift SN to date ($z\sim2$). From SLSN~I data only, we find $\Omega_{\rm M}=0.38^{+0.24}_{-0.19}$ (stat+sys) and an $\mathrm{rms}=0.27$\,mag. We also explored a $w_0w_a$CDM cosmological model combining our SLSN~I sample with the JLA sample and measurements from the CMB, finding that only a small improvement can be made in the constraints on $w_0$ and $w_a$ by 4\% in terms of their FoM.  
We have also simulated future datasets, and demonstrated their potential to reduce the current statistical uncertainties by a factor of ten on $\Omega_{\rm M}$, making them comparable to those found using current SN Ia samples. The FoM of the CMB+WMAP+JLA+SLSNe set up will increase, providing an improvement of a factor 2-3 in the precision of cosmological constraints and also offering a longer redshift baseline. This represents a proof-of-concept of the current potential and future strengths of SLSN~I in cosmology. The key output of this study is that it empowers the investigation of the behaviour of our Universe ($\Omega_{\rm M}$, $w_0$, $w_a$) up to redshifts that cannot be explored using other SNe from the ground ($z>1.5$). 

\section*{Acknowledgements}

This paper has gone through internal review by the DES
collaboration. It has Fermilab Preprint number 19-115-AE and DES publication number 13387.
We acknowledge support from EU/FP7-ERC grant 615929.
R.C.N. would like to acknowledge support from STFC grant
ST/N000688/1 and the Faculty of Technology at the
University of Portsmouth. 
L.G. was funded by the European Union's Horizon 2020 research and innovation programme under the Marie Sk\l{}odowska-Curie grant agreement No. 839090. This work has been partially supported by the Spanish grant PGC2018-095317-B-C21 within the European Funds for Regional Development (FEDER).
Funding for the DES Projects has been provided by the U.S. Department of Energy, the U.S. National Science Foundation, the Ministry of Science and Education of Spain, 
the Science and Technology Facilities Council of the United Kingdom, the Higher Education Funding Council for England, the National Center for Supercomputing 
Applications at the University of Illinois at Urbana-Champaign, the Kavli Institute of Cosmological Physics at the University of Chicago, 
the Center for Cosmology and Astro-Particle Physics at the Ohio State University,
the Mitchell Institute for Fundamental Physics and Astronomy at Texas A\&M University, Financiadora de Estudos e Projetos, 
Funda{\c c}{\~a}o Carlos Chagas Filho de Amparo {\`a} Pesquisa do Estado do Rio de Janeiro, Conselho Nacional de Desenvolvimento Cient{\'i}fico e Tecnol{\'o}gico and 
the Minist{\'e}rio da Ci{\^e}ncia, Tecnologia e Inova{\c c}{\~a}o, the Deutsche Forschungsgemeinschaft and the Collaborating Institutions in the Dark Energy Survey. 
The Collaborating Institutions are Argonne National Laboratory, the University of California at Santa Cruz, the University of Cambridge, Centro de Investigaciones Energ{\'e}ticas, 
Medioambientales y Tecnol{\'o}gicas-Madrid, the University of Chicago, University College London, the DES-Brazil Consortium, the University of Edinburgh, 
the Eidgen{\"o}ssische Technische Hochschule (ETH) Z{\"u}rich, 
Fermi National Accelerator Laboratory, the University of Illinois at Urbana-Champaign, the Institut de Ci{\`e}ncies de l'Espai (IEEC/CSIC), 
the Institut de F{\'i}sica d'Altes Energies, Lawrence Berkeley National Laboratory, the Ludwig-Maximilians Universit{\"a}t M{\"u}nchen and the associated Excellence Cluster Universe, 
the University of Michigan, the National Optical Astronomy Observatory, the University of Nottingham, The Ohio State University, the University of Pennsylvania, the University of Portsmouth, 
SLAC National Accelerator Laboratory, Stanford University, the University of Sussex, Texas A\&M University, and the OzDES Membership Consortium.
Based in part on observations at Cerro Tololo Inter-American Observatory, National Optical Astronomy Observatory, which is operated by the Association of 
Universities for Research in Astronomy (AURA) under a cooperative agreement with the National Science Foundation.
The DES data management system is supported by the National Science Foundation under Grant Numbers AST-1138766 and AST-1536171.
The DES participants from Spanish institutions are partially supported by MINECO under grants AYA2015-71825, ESP2015-66861, FPA2015-68048, SEV-2016-0588, SEV-2016-0597, and MDM-2015-0509, 
some of which include ERDF funds from the European Union. IFAE is partially funded by the CERCA program of the Generalitat de Catalunya.
Research leading to these results has received funding from the European Research
Council under the European Union's Seventh Framework Program (FP7/2007-2013) including ERC grant agreements 240672, 291329, and 306478.
We  acknowledge support from the Australian Research Council Centre of Excellence for All-sky Astrophysics (CAASTRO), through project number CE110001020, and the Brazilian Instituto Nacional de Ci\^encia
e Tecnologia (INCT) e-Universe (CNPq grant 465376/2014-2).

This manuscript has been authored by Fermi Research Alliance, LLC under Contract No. DE-AC02-07CH11359 with the U.S. Department of Energy, Office of Science, Office of High Energy Physics. The United States Government retains and the publisher, by accepting the article for publication, acknowledges that the United States Government retains a non-exclusive, paid-up, irrevocable, world-wide license to publish or reproduce the published form of this manuscript, or allow others to do so, for United States Government purposes.


\section*{Affiliations}
\noindent
{\it
$^{1}$ School of Physics \& Astronomy, Cardiff University, Queens Buildings, The Parade, Cardiff, CF24 3AA, UK\\
$^{2}$ School of Physics and Astronomy, University of Southampton, Southampton, SO17 1BJ, UK\\
$^{3}$ DARK, Niels Bohr Institute, University of Copenhagen, Lyngbyvej 2, DK-2100 Copenhagen $\O$, Denmark\\
$^{4}$ Department of Physics and Astronomy, University of North Georgia, GA 30597, USA\\
$^{5}$ Institute of Cosmology and Gravitation, University of Portsmouth, Portsmouth PO1 3FX, UK\\
$^{6}$ Universite Clermont Auvergne, CNRS/IN2P3, LPC, F-63000 Clermont-Ferrand, France\\
$^{7}$  Department of Physics and Astronomy, University of Pennsylvania, Philadelphia, PA 19104, USA\\
$^{8}$  George P. and Cynthia Woods Mitchell Institute for Fundamental Physics and Astronomy, and Department of Physics and Astronomy, Texas A\&M University, College Station, TX 77843,  USA\\
$^{9}$  School of Mathematics and Physics, University of Queensland,  Brisbane, QLD 4072, Australia\\
$^{10}$  Departamento de F\'isica Te\'orica y del Cosmos, Universidad de Granada, E-18071 Granada, Spain\\
$^{11}$  Department of Astronomy and Astrophysics, University of Chicago, Chicago, IL 60637, USA\\
$^{12}$  Kavli Institute for Cosmological Physics, University of Chicago, Chicago, IL 60637, USA\\
$^{13}$  Lawrence Berkeley National Lab, 1 Cyclotron Rd., Berkeley, CA, 94720\\
$^{14}$  Division of Science, National Astronomical Observatory of Japan, 2-21-1 Osawa, Mitaka, Tokyo 181-8588, Japan\\
$^{15}$  Cerro Tololo Inter-American Observatory, National Optical Astronomy Observatory, Casilla 603, La Serena, Chile\\
$^{16}$  Fermi National Accelerator Laboratory, P. O. Box 500, Batavia, IL 60510, USA\\
$^{17}$  Instituto de Fisica Teorica UAM/CSIC, Universidad Autonoma de Madrid, 28049 Madrid, Spain\\
$^{18}$  CNRS, UMR 7095, Institut d'Astrophysique de Paris, F-75014, Paris, France\\
$^{19}$  Sorbonne Universit\'es, UPMC Univ Paris 06, UMR 7095, Institut d'Astrophysique de Paris, F-75014, Paris, France\\
$^{20}$  Department of Physics \& Astronomy, University College London, Gower Street, London, WC1E 6BT, UK\\
$^{21}$  Kavli Institute for Particle Astrophysics \& Cosmology, P. O. Box 2450, Stanford University, Stanford, CA 94305, USA\\
$^{22}$  SLAC National Accelerator Laboratory, Menlo Park, CA 94025, USA\\
$^{23}$  Centro de Investigaciones Energ\'eticas, Medioambientales y Tecnol\'ogicas (CIEMAT), Madrid, Spain\\
$^{24}$  Laborat\'orio Interinstitucional de e-Astronomia - LIneA, Rua Gal. Jos\'e Cristino 77, Rio de Janeiro, RJ - 20921-400, Brazil\\
$^{25}$  Department of Astronomy, University of Illinois at Urbana-Champaign, 1002 W. Green Street, Urbana, IL 61801, USA\\
$^{26}$  National Center for Supercomputing Applications, 1205 West Clark St., Urbana, IL 61801, USA\\
$^{27}$  Institut de F\'{\i}sica d'Altes Energies (IFAE), The Barcelona Institute of Science and Technology, Campus UAB, 08193 Bellaterra (Barcelona) Spain\\
$^{28}$  Institut d'Estudis Espacials de Catalunya (IEEC), 08034 Barcelona, Spain\\
$^{29}$  Institute of Space Sciences (ICE, CSIC),  Campus UAB, Carrer de Can Magrans, s/n,  08193 Barcelona, Spain\\
$^{30}$  Physics Department, 2320 Chamberlin Hall, University of Wisconsin-Madison, 1150 University Avenue Madison, WI  53706-1390\\
$^{31}$  Department of Physics, IIT Hyderabad, Kandi, Telangana 502285, India\\
$^{32}$  Department of Astronomy/Steward Observatory, University of Arizona, 933 North Cherry Avenue, Tucson, AZ 85721-0065, USA\\
$^{33}$  Jet Propulsion Laboratory, California Institute of Technology, 4800 Oak Grove Dr., Pasadena, CA 91109, USA\\
$^{34}$  Department of Astronomy, University of Michigan, Ann Arbor, MI 48109, USA\\
$^{35}$  Department of Physics, University of Michigan, Ann Arbor, MI 48109, USA\\
$^{36}$  Institute of Astronomy, University of Cambridge, Madingley Road, Cambridge CB3 0HA, UK\\
$^{37}$  Kavli Institute for Cosmology, University of Cambridge, Madingley Road, Cambridge CB3 0HA, UK\\
$^{38}$  Department of Physics, Stanford University, 382 Via Pueblo Mall, Stanford, CA 94305, USA\\
$^{39}$  Observat\'orio Nacional, Rua Gal. Jos\'e Cristino 77, Rio de Janeiro, RJ - 20921-400, Brazil\\
$^{40}$  Santa Cruz Institute for Particle Physics, Santa Cruz, CA 95064, USA\\
$^{41}$  Center for Cosmology and Astro-Particle Physics, The Ohio State University, Columbus, OH 43210, USA\\
$^{42}$  ASTRAVEO LLC, PO Box 1668, MA 01931\\
$^{43}$  Australian Astronomical Optics, Macquarie University, North Ryde, NSW 2113, Australia\\
$^{44}$  The Research School of Astronomy and Astrophysics, Australian National University, Canberra, ACT 2611, Australia\\
$^{45}$  Departamento de F\'isica Matem\'atica, Instituto de F\'isica, Universidade de S\~ao Paulo, CP 66318, S\~ao Paulo, SP, 05314-970, Brazil\\
$^{46}$  Instituci\'o Catalana de Recerca i Estudis Avan\c{c}ats, E-08010 Barcelona, Spain\\
$^{47}$  Department of Astrophysical Sciences, Princeton University, Peyton Hall, Princeton, NJ 08544, USA\\
$^{48}$ Department of Physics and Astronomy, Pevensey Building, University of Sussex, Brighton, BN1 9QH, UK\\
$^{49}$  Brandeis University, Physics Department, 415 South Street, Waltham MA 02453\\
$^{50}$  Instituto de F\'isica Gleb Wataghin, Universidade Estadual de Campinas, 13083-859, Campinas, SP, Brazil\\
$^{51}$  Computer Science and Mathematics Division, Oak Ridge National Laboratory, Oak Ridge, TN 37831\\
$^{52}$  Argonne National Laboratory, 9700 South Cass Avenue, Lemont, IL 60439, USA\\
$^{53}$  Korea Astronomy and Space Science Institute, Yuseong-gu, Daejeon, 305-348, Korea\\
$^{54}$  School of Mathematics and Physics, University of Queensland,  Brisbane, QLD 4072, Australia\\
$^{55}$  INAF, Astrophysical Observatory of Turin, I-10025 Pino Torinese, Italy\\
$^{56}$  Centre for Astrophysics \& Supercomputing, Swinburne University of Technology, Victoria 3122, Australia\\
$^{57}$  Sydney Institute for Astronomy, School of Physics, A28, The University of Sydney, NSW 2006, Australia\\
$^{58}$ Université de Lyon 1, CNRS/IN2P3, Institut de Physique des Deux Infinis, 69622, Villeurbanne Cedex, France\\
$^{59}$ Finnish Centre for Astronomy with ESO (FINCA), FI-20014 University of Turku, Finland\\
$^{60}$ Tuorla Observatory, Department of Physics and Astronomy, FI-20014 University of Turku, Finland
}

\section*{Data availability}
The data underlying this article are available in the article and in its online supplementary material. The 4OPS code is available at the following GitHub account \hyperlink{https://github.com/cinserra/4OPS}{https://github.com/cinserra/4OPS}. Any data not presented in the above table or references will be shared on reasonable request to the corresponding author.



\bibliographystyle{mnras}
\bibliography{SLSN_cosmology_arxiv.bib} 

\appendix
\section{4OPS}\label{app:4OPS}
\begin{figure*}
\center
\includegraphics[width=16cm]{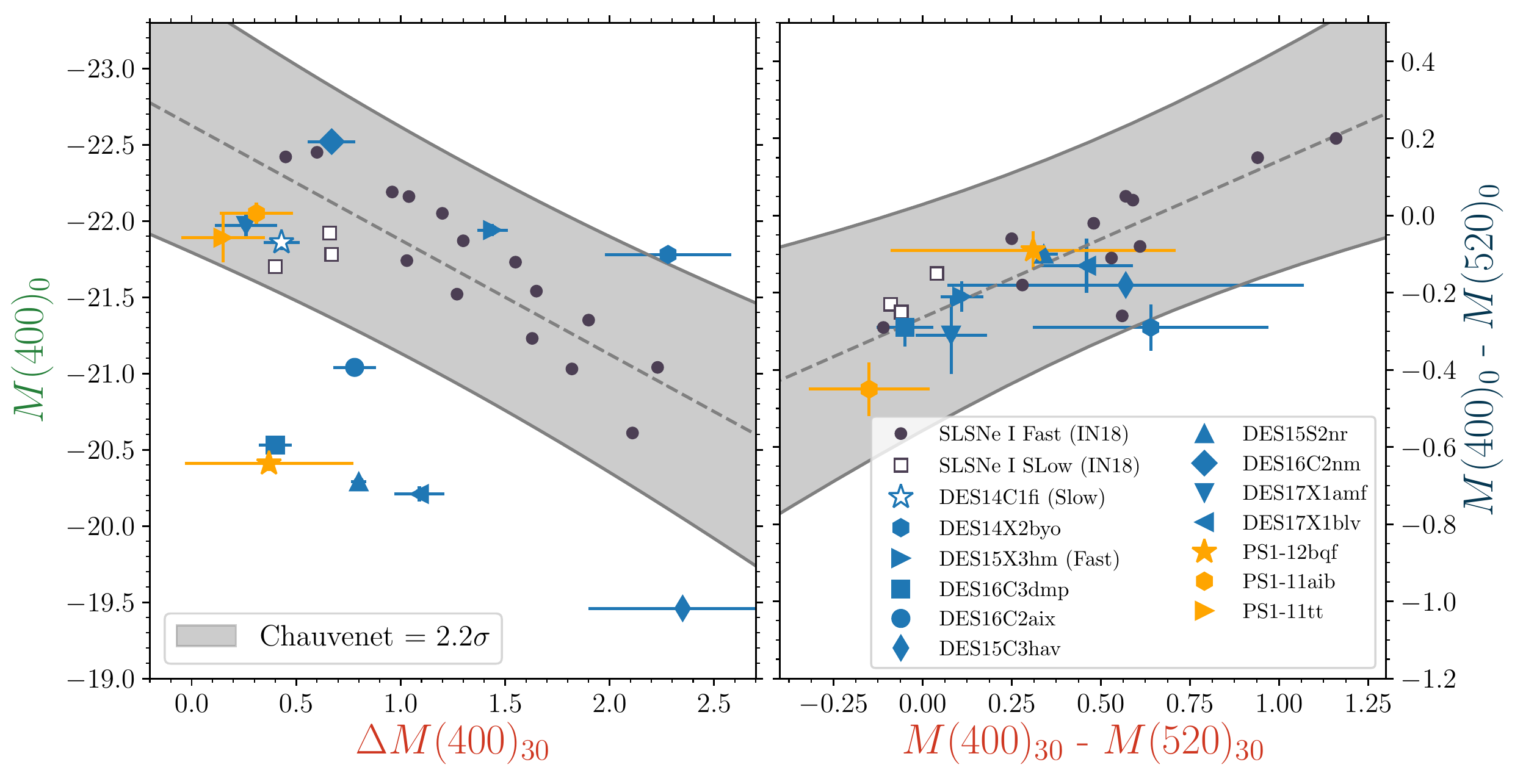}
\caption{A reduced version of the Four Observables Parameter Space (4OPS) for SLSNe~I. Data are taken from DES, literature, and other surveys' sample papers, that made our first two quality cuts. The left panel shows the magnitude at peak, vs the decline rate over 30 days past peak.  The right panel shows the colour at peak vs the colour at 30 days post peak. The literature objects, both Fast (circles) and Slow (open squares) are shown together to their best fit of the weighted linear regression (dashed, black line) and with the $2.2\sigma$ confidence bands defined by the Chauvenet's criterion. We also include DES16C2nm, which was not presented before in the literature sample \citep{in18c}. }
\label{fig:4OPS_orig}
\end{figure*}





\bsp	
\label{lastpage}
\end{document}